%% file: main.tex
\PassOptionsToPackage{table}{xcolor}

\documentclass[acmsmall,screen,nonacm]{acmart}

\setcopyright{none}
\graphicspath{{figures/}} 

\usepackage{float}
\floatstyle{plaintop}
\restylefloat{table}
\include{listingsconfig}

\include{db}
\usepackage{subcaption}

\usepackage{tikzit}
\input{tree.tikzstyles}

\usepackage{mathpartir}
\usepackage{stmaryrd}
\usepackage{geometry}
\usepackage{mdframed}
\usepackage[tight-spacing=true]{siunitx}
\usepackage{wrapfig}
\usepackage{xspace}

\usepackage{enumitem}
\setitemize{noitemsep,topsep=0pt,parsep=0pt,partopsep=0pt}
\setenumerate{noitemsep,topsep=0pt,parsep=0pt,partopsep=0pt}

\mathchardef\mhyphen="2D 

\newcommand{\keyw}[1]{\ensuremath{\mathsf{#1}\xspace}}

\theoremstyle{remark}

\newcommand{\link}[3]{\href{https://github.com/bagnalla/zar/tree/release-pldi23/#1.v\#L#2}{#3}}

\newcommand{\eR}{\mathbb{R}^\infty_{\geq 0}}
\newcommand{\beR}{\mathbb{R}^{\le 1}_{\geq 0}}
\newcommand{\cftree}{\mathcal T^{\mathit{cf}}}
\newcommand{\itree}[1]{\mathcal T^{\mathit{it}}_{#1}}

\newcommand{\bind}[1]{\gg\!\!=_{\mathsf{#1}}}

\newcommand{\cons}[2]{\mathsf{\mathbf{#1}^{#2}}}

\newcommand{\prop}{\mathbb{P}}
\newcommand{\pred}[1]{#1 \rightarrow \prop}

\newcommand{\unit}{\mathbf{1}}
\newcommand{\true}{\mathbf{true}}
\newcommand{\false}{\mathbf{false}}
\renewcommand{\tt}{\mathsf{()}}
\newcommand{\bool}{\mathbb{B}}
\newcommand{\nat}{\mathbb{N}}
\newcommand{\cpGCL}{\mathsf{cpGCL}}
\newcommand{\expr}{\Sigma \rightarrow \textsf{val}}
\newcommand{\expectation}{\Sigma \rightarrow \eR}
\newcommand{\bexpectation}{\Sigma \rightarrow \beR}
\renewcommand{\wp}{\mathsf{wp}}
\newcommand{\wlp}{\mathsf{wlp}}
\newcommand{\cwp}{\mathsf{cwp}}
\newcommand{\twp}{\mathsf{twp}}

\newcommand{\tcwp}{\mathsf{tcwp}}
\newcommand{\flip}[2]{\mathsf{flip} \: #1 \: #2}

\newcommand{\assign}[2]{#1 \leftarrow #2}
\newcommand{\seq}[2]{#1 ; \: #2}
\newcommand{\obs}[1]{\cons{observe}{} \: #1}
\newcommand{\ite}[3]{\cons{if}{} \: #1 \: \cons{then}{} \: #2 \: \cons{else}{} \: #3}
\newcommand{\choice}[3]{\{ \: #2 \: \} \: [#1] \: \{ \: #3 \: \}}
\newcommand{\uniform}[2]{\cons{uniform}{} \: #1 \: #2}
\newcommand{\while}[2]{\cons{while}{} \: #1 \: \cons{do}{} \: #2 \: \cons{end}{}}

\newcommand{\Iid}{i.i.d.}
\newcommand{\iid}{i.i.d. }
\newcommand{\zar}{\keyw{Zar}\xspace}

\begin{document}
\title{Formally Verified Samplers from Probabilistic Programs with Loops and Conditioning}

\thanks{This is an extended version of the paper appearing in PLDI'23.}

\author{Alexander Bagnall}
\email{abagnalla@gmail.com}
\affiliation{
  \institution{Ohio University}
  \city{Athens}
  \state{Ohio}
  \country{USA}}
\orcid{0000-0001-6593-0661}
\author{Gordon Stewart}
\affiliation{
  \institution{BedRock Systems, Inc.}
  \city{Boston}
  \state{Massachusetts}
  \country{USA}}
\email{gordon@bedrocksystems.com}
\orcid{0000-0003-0244-2980}
\author{Anindya Banerjee}
\affiliation{
  \institution{IMDEA Software Institute}
  \city{Pozuelo de Alarcon}
  \state{Madrid}
  \country{Spain}}
\email{anindya.banerjee@imdea.org}
\orcid{0000-0001-9979-1292}

\authorsaddresses{} 

\begin{abstract}

We present \zar: a formally verified compiler pipeline from discrete
probabilistic programs with unbounded loops in the conditional
probabilistic guarded command language ($\cpGCL$) to proved-correct
executable samplers in the random bit model. We exploit the key idea
that all discrete probability distributions can be reduced to unbiased
coin-flipping schemes. The compiler pipeline first translates $\cpGCL$
programs into \textit{choice-fix} trees, an intermediate
representation suitable for reduction of biased probabilistic
choices. Choice-fix trees are then translated to coinductive
interaction trees for execution within the random bit model. The
correctness of the composed translations establishes
the \textit{sampling equidistribution theorem}: compiled samplers are
correct wrt.~the conditional weakest pre-expectation semantics of
$\cpGCL$ source programs. \zar is implemented and fully verified in
the Coq proof assistant. We extract verified samplers to OCaml and
Python and empirically validate them on a number of illustrative
examples.

\end{abstract}




\maketitle

\section{Introduction}
\label{sec:introduction}

Probabilistic programming languages~\cite{DBLP:conf/icse/GordonHNR14,
bingham2019pyro, goodman2012church} formalize probabilistic systems by
modeling them as programs with random sampling and
conditioning. Unlike conventional programs, for which meaning is
deduced from executions over states or sets of states, probabilistic
programs are defined by their posterior distributions for given
inputs. Calculating this posterior distribution is
called \textit{inference}. In cases in which it is infeasible to
calculate the posterior directly, probabilistic programming languages
(PPLs) typically support sampling from this distribution. Many
standard semantic notions such as weakest precondition transformers
have analogues -- e.g., weakest pre-expectation transformers -- in
PPLs.

Inference on probabilistic programs (PPs) is automated by compiling
the programs to Markov Chain Monte Carlo (MCMC)
samplers~\cite{HuangTM17} or to other specialized
representations~\cite{holtzen2019symbolic, holtzen2020scaling}.
Similarly, systems like Pyro~\cite{bingham2019pyro} use techniques for
semi-automated inference.  Automation helps separate concerns: the
programmer specifies a probabilistic model in a convenient high-level
language, and the inference engine takes care of the details of
calculating the posterior distribution.  At the same time, errors in
execution models of probabilistic systems are especially difficult to
detect and diagnose~\cite{dutta2018testing, dutta2019storm}, and
attempts at empirical validation may fail to detect small biases and
low-probability error conditions. The standard belief propagation
algorithm for inference may converge to the wrong equilibrium or fail
to converge at all~\cite{yedidia2003understanding}, and MCMC samplers
may falsely appear to have converged to the desired stationary
distribution (known as
``pseudo-convergence'')~\cite{geyer2011introduction}.  Even the
straightforward task of uniform sampling is notoriously susceptible to
``modulo bias''~\cite{kudelski2020modulo}, leading to violations of
cryptographic guarantees~\cite{aranha2020ladderleak} due to improper
use of the modulus operator to restrict the range of the uniform
distribution.

We implement \zar: a formally verified compiler from the conditional
probabilistic guarded command language
($\cpGCL$~\cite{olmedo2018conditioning}) to proved-correct samplers in
the random bit model~\cite{saad2020sampling}, in which samplers are
provided a stream of independent and identically distributed (\Iid)
random bits drawn from a \textit{uniform} distribution. Samplers
generated by \zar are guaranteed, under reasonable assumptions about
the source of randomness (Section~\ref{subsec:equidistribution}), to
produce samples from the true posterior of their source programs, and
thus provide a foundation for high-assurance sampling and Monte
Carlo-based~\cite{rubinstein2016simulation} inference. Additionally
(Section~\ref{subsec:uniform-sampling}), we apply the \zar compiler
backend to verify samplers for discrete uniform
distributions. \href{https://github.com/bagnalla/zar/tree/release-pldi23}{\zar}
is implemented and fully verified in the Coq proof assistant.


\subsection{Challenges}

\lstset{numbers=left}

\begin{figure}[t]
  \centering
  \begin{subfigure}{.4\textwidth}
    \centering
      \begin{lstlisting}[language=cpGCL,mathescape=true]
primes ($p : \mathbb{Q}$) :=
  { $b$ $\leftarrow$ $\true$ } [$p$] { $b$ $\leftarrow$ $\false$ };
  while $b$ do
    $h$ $\leftarrow$ $h + 1$;
    { $b$ $\leftarrow$ $\true$ } [$p$] { $b$ $\leftarrow$ $\false$ }
  end;
  observe $h$ is prime
      \end{lstlisting}

    \caption{`$\mathsf{primes}$' $\cpGCL$ program with geometric
    posterior over the prime numbers.}

    \label{prog:intro-primes}
  \end{subfigure}%
  \begin{subfigure}{.5\textwidth}
    \centering
    \includegraphics[width=1.25\linewidth]{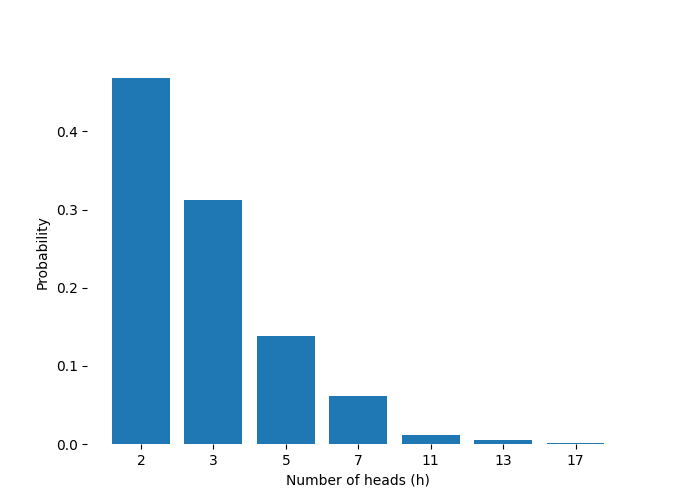}

    \caption{True posterior over $h$ with $p=\frac{2}{3}$.}

    \label{fig:intro-primes-bar}
  \end{subfigure}%

  \caption{Geometric primes program (left) and its posterior
  distribution over $h$ (right).}

  \label{fig:intro-primes}
\end{figure}

\lstset{numbers=none}

To understand the challenges, consider the `$\mathsf{primes}$'
$\cpGCL$ program in Figure~\ref{prog:intro-primes}, which computes a
geometric posterior over the prime numbers as shown in
Figure~\ref{fig:intro-primes-bar}.  This program combines three
fundamental features complicating inference: 1) nonuniform (biased)
probabilistic choice, 2) unbounded loop-carried dataflow (a
``non-\Iid'' loop~\cite{kaminski2019advanced}), and 3)
conditioning. The variable $b$ is drawn from a Bernoulli distribution
with probability $p$ of ``heads'' (lines $2$, $5$). The variable $h$
(with initial value $0$ and updated on line $4$) counts the number of
heads encountered before flipping tails. Finally, the terminal program
state is conditioned on $h$ being prime (line $7$).

\paragraph*{Eliminating Bias}
The probabilistic choices in the program of
Figure~\ref{prog:intro-primes} are specialized in
Figure~\ref{fig:intro-primes-bar} to the nonuniform bias $p
= \frac{2}{3}$. To obtain a sampler in the random bit model, the
program must be transformed into a semantically equivalent one in
which all choices have bias $\frac{1}{2}$. Moreover, probability
expressions in $\cpGCL$ can be functions of the program state, so
reduction of biased choices is not always possible via direct
source-to-source translation (e.g., the probability expression on line
$5$ could depend on variable $h$). To address this state dependence
and the use of nonuniform biases, we develop a new intermediate
representation called choice-fix trees
(Section~\ref{subsec:compiling-to-cf-trees}). We compile $\cpGCL$
programs to the choice-fix representation and debias choice-fix trees
to generate samplers in the random bit model.

\paragraph*{Unbounded and Non-\iid Loops}
The loop in Figure~\ref{prog:intro-primes} is unbounded; it is not
guaranteed to terminate within any fixed number of iterations, and can
diverge (though with probability 0) when only heads are flipped. The
tasks of sampling and inference are greatly complicated by the
infinitary nature of unbounded loops, and thus much previous work on
discrete PPs is limited to bounded
loops~\cite{chavira2008probabilistic, HuangTM17, holtzen2019symbolic,
holtzen2020scaling}. Formal reasoning about infinitary computations
requires substantial use
of \textit{coinduction}~\cite{kozen2017practical}, which is
notoriously difficult to use in proof assistants like
Coq~\cite{hur2013power}.

Moreover, the loop in Figure~\ref{prog:intro-primes} is ``non-\Iid'';
the update of counter variable $h$ on line $4$ induces nontrivial data
dependence between iterations of the loop, and consequently every
value of $h \ge 0$ occurs with nonzero probability (the posterior
has \textit{infinite support}). Many interesting probabilistic
programs such as the discrete Gaussian (see
Section~\ref{subsec:hare-and-tortoise}) exhibit such ``loop-carried
dependence''~\cite{allen1987automatic}. Prior work on automated
inference of unbounded loops and conditioning on observations in
probabilistic programs has been restricted to the subclass of \iid
loops, i.e., those without loop-carried
dependence~\cite{bagnall1coinductive}.

\zar compiles the `$\mathsf{primes}$' program to an
executable \textit{interaction tree} (ITree)~\cite{xia2019interaction}
formally guaranteed to produce samples from the geometric posterior
shown in Fig~\ref{fig:intro-primes-bar} when provided uniform random
bits from its environment (see Section~\ref{subsec:geometric-primes}
for empirical evaluation). The coinductive type of ITrees, while
suitable for encoding potentially unbounded processes, is deceptively
difficult to reason about formally. Coq's built-in mechanism for
coinduction is often not
sufficient~\cite{chlipala2022certified,hur2013power}. To facilitate
reasoning on coinductive representations of samplers, we employ
concepts from domain theory such as
Scott-continuity~\cite{abramsky1994domain} and algebraic
CPOs~\cite{gunter1992semantics} (see related discussion in
Section~\ref{subsec:generating-itrees}).

\paragraph*{Correctness of Samplers}
Verified compilers of conventional programming languages like C have
somewhat well understood correctness guarantees (though
see~\cite{patterson2019next}). CompCert~\cite{leroy2009formal}, for
example, uses a simulation argument to prove a form of behavioral
equivalence of source and target programs.  Writing the specification
of a compiler for a PPL is less straightforward. What does
``behavioral equivalence'' even mean when the result of the
compilation pipeline is a probabilistic sampler that depends on a
source of randomness?

A key idea of this paper is that the proof of a PPL compiler is
essentially a reduction: as input, it takes a source of randomness (in
our case, uniformly distributed random bits) and as output it produces
a sampler on the posterior distribution generated by the conditional weakest
pre-expectation semantics ($\cwp$) of the program being compiled. We thus reduce the
problem of sampling a program's posterior distribution to the comparatively simpler
problem of sampling uniformly random bits. Making this
reduction work formally means precisely characterizing the input
source of randomness and the distributional correctness of the output
sampler. We specify the input randomness in Section
\ref{subsec:source-of-randomness}, drawing on the classic theory of
uniform distribution modulo 1~\cite{weyl1916gleichverteilung}. We
characterize distributional correctness by proving that our samplers
satisfy an \textit{equidistribution theorem}
(Section~\ref{sec:correctness-of-sampling}) wrt. the $\cwp$ semantics
of source programs.

\subsection{Contributions}
\noindent\textbf{Concept.}
We implement \zar: a formally verified compilation pipeline from
discrete probabilistic programs with conditioning to proved-correct
executable samplers in the random bit model, exploiting the key idea
that all discrete distributions can be reduced to unbiased
coin-flipping schemes~\cite{knuth1976complexity}, and culminating in
the \textit{sampling equidistribution} theorem
(Theorem~\ref{theorem:cpGCL-equidistribution}) establishing
correctness of compiled samplers. The entire system is fully
implemented and verified in Coq.

\noindent\textbf{Technical.}
The \zar system includes:
\begin{itemize}
  \item[-] a formalization of $\cpGCL$ and its associated $\cwp$
semantics (Section~\ref{sec:syntax-semantics-cpGCL}),

  \item[-] an intermediate representation for $\cpGCL$ programs
  called \textit{choice-fix} trees (Section~\ref{subsec:cf-trees}),
  enabling optimizations and essential program transformations (e.g.,
  elimination of redundant choices and reduction to the random bit
  model),

  \item[-] a compiler pipeline
(Section~\ref{subsec:compiling-to-cf-trees}) from $\cpGCL$ to ITree
samplers (Section~\ref{subsec:generating-itrees}),

  \item[-] statement and proof of a general result establishing the
  correctness of compiled samplers wrt.~the $\cwp$ semantics of source
  programs, based on the notion of equidistribution
  (Section~\ref{sec:correctness-of-sampling}), and

  \item[-] a Python 3 package for high-assurance uniform
  sampling (Section~\ref{subsec:uniform-sampling}) as a thin wrapper
  around proved-correct samplers extracted from Coq.
\end{itemize}

\begin{figure}[t]
  \centering
  \includegraphics[width=\textwidth]{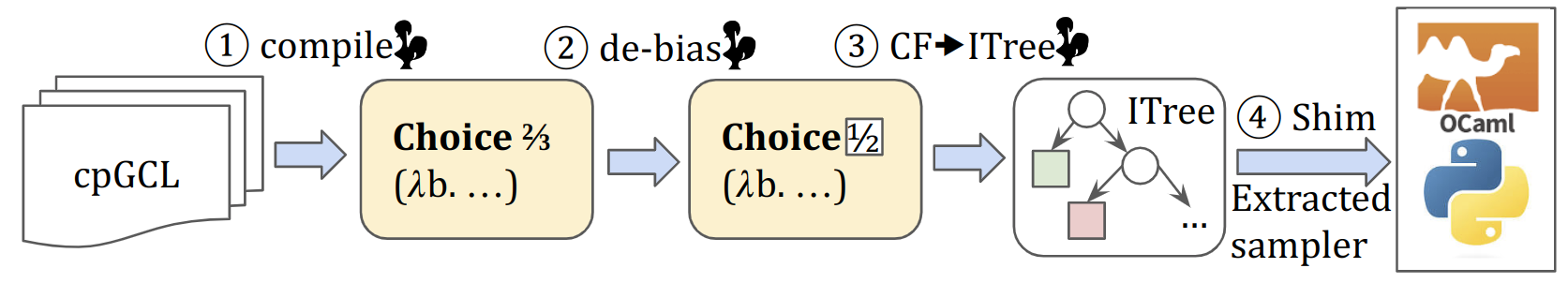}
  
  \caption{\zar pipeline diagram showing (1) the compiler from
  $\cpGCL$ to CF trees (Section~\ref{subsec:cf-trees}), (2) debiasing
  of probabilistic choices
  (Section~\ref{subsec:compiling-to-cf-trees}), (3) generation of
  interaction trees from CF trees
  (Section~\ref{subsec:generating-itrees}), and (4) extraction for
  efficient execution in OCaml and Python
  (Section~\ref{sec:empirical-validation}).}

  \label{fig:system-architecture}
\end{figure}


\noindent{\textbf{Evaluation.} We perform empirical validation of illustrative
examples (Section~\ref{sec:empirical-validation}) including posterior
inference over a simulated race between a hare and tortoise
(Section~\ref{subsec:hare-and-tortoise}, inspired by~\cite[Section
1]{SzymcakK20}).

\noindent{\textbf{Source Code.}} Embedded hyperlinks in the PDF point to the
underlying \href{https://github.com/bagnalla/zar/blob/release-pldi23}{Coq
sources} The Python 3 package for uniform sampling
is \href{https://github.com/bagnalla/zar/tree/release-pldi23/python/zar}{available
in the \zar repository}.

\noindent{\textbf{Axiomatic Base.} We extend the type theory of Coq with
excluded middle, indefinite description, and functional
extensionality~\cite{chargueraud2017axioms}. We also use Coq's
standard real number library and a
custom \link{cotree}{\cwpcotreezext}{extensionality axiom} for
coinductive trees.

\subsection{Current Limitations}

\zar
supports only discrete probabilistic $\cpGCL$ programs (which are
naturally suited for many applications~\cite{holtzen2020scaling}) that
terminate either absolutely or almost surely (i.e., with probability
$1$). Probabilities appearing in $\cpGCL$ programs must be rational
numbers. We provide no guarantees regarding time/space or entropy
usage (number of random bits required to obtain a sample), although we
observe near entropy-optimality in some cases
(cf.~Section~\ref{subsec:uniform-sampling}). We verify only
the \textit{compiler pipeline}. Verification of $\cpGCL$ programs
using a program logic that is sound wrt. the $\cwp$ semantics is
beyond the scope of this paper. Proofs on $\cpGCL$ programs wrt. their
$\cwp$ semantics can, however, be composed with our compiler
correctness proofs
(Theorems~\ref{theorem:cf-tree-compiler-correctness},~\ref{theorem:cpGCL-equidistribution})
to obtain end-to-end guarantees on generated samplers. In future work,
we plan to address the current limitation to discrete programs by
extending $\cpGCL$ with sampling from continuous random variates over
the unit interval. To address entropy use, we have plans to compile to
MCMC-based sampling processes.




\section{Syntax and Semantics of $\cpGCL$}
\label{sec:syntax-semantics-cpGCL}

This section presents $\cpGCL$ together with its conditional weakest
pre-expectation semantics, $\cwp$~\cite{olmedo2018conditioning}. We
extend $\cpGCL$ and $\cwp$ as follows:

\begin{enumerate}

\item We add to $\cpGCL$ an additional command for drawing samples uniformly at random from a range of natural numbers (Section~\ref{subsec:cpGCL}).

\item We let probability expressions appearing in choice commands depend on the program state
  (Section~\ref{subsec:cpGCL}).

\item We extend the weakest (liberal) expectation transformers $\wp$ ($\wlp$) to better support reasoning about the probability of observation failure (Section~\ref{subsec:cwp}).

\end{enumerate}

\subsection{Syntax of $\cpGCL$}
\label{subsec:cpGCL}

\begin{definition}[\link{cpGCL}{\cwpcpGCL}{$\cpGCL$}]
\label{def:cpGCLsyn}
  Type $\cpGCL$ is defined inductively as:
  \begin{mathpar}
    \inferrule [\link{cpGCL}{\cwpCSkip}{\textsf{cpGCL-skip}}]
               { }
               { \cons{skip}{} : \cpGCL }
    \and
    \inferrule [\link{cpGCL}{\cwpCAssign}{\textsf{cpGCL-assign}}]
               { x : \mathsf{ident} \\
                 e : \expr }
               { \assign{x}{e} : \cpGCL }
    \and
    \inferrule [\link{cpGCL}{\cwpCSeq}{\textsf{cpGCL-seq}}]
               { c_1 : \cpGCL \\
                 c_2 : \cpGCL }
               { \seq{c_1}{c_2} : \cpGCL }
    \and
    \inferrule [\link{cpGCL}{\cwpCObserve}{\textsf{cpGCL-observe}}]
               { e : \Sigma \rightarrow \bool }
               { \obs{e} : \cpGCL }
    \and
    \inferrule [\link{cpGCL}{\cwpCIte}{\textsf{cpGCL-ite}}]
               { e : \Sigma \rightarrow \bool \\
                 c_1 : \cpGCL \\
                 c_2 : \cpGCL }
               { \ite{e}{c_1}{c_2} : \cpGCL }
    \and
    \inferrule [\link{cpGCL}{\cwpCChoice}{\textsf{cpGCL-choice}}]
               { p : \Sigma \rightarrow \mathbb{Q} \\
                 \forall \sigma : \Sigma, \: 0 \le p \: \sigma \le 1 \\
                 c_1 : \cpGCL \\
                 c_2 : \cpGCL }
               { \choice{p}{c_1}{c_2} : \cpGCL }
    \and
    \inferrule [\link{cpGCL}{\cwpCUniform}{\textsf{cpGCL-uniform}}]
               { e : \Sigma \rightarrow \nat \\
                 \forall \sigma : \Sigma, \: 0 < e \: \sigma \\
                 k : \nat \rightarrow \cpGCL }
               { \uniform{e}{k} : \cpGCL }
    \and
    \inferrule [\link{cpGCL}{\cwpCWhile}{\textsf{cpGCL-while}}]
               { e : \Sigma \rightarrow \bool \\
                 c : \cpGCL }
               { \while{e}{c} : \cpGCL }
  \end{mathpar}
\end{definition}

\noindent $\cpGCL$ (Definition~\ref{def:cpGCLsyn}) extends the guarded command
language~\cite{dijkstra1975guarded} with:

\begin{enumerate}
  \item \textbf{Probabilistic choice:} Given expression $e
    : \Sigma \rightarrow \mathbb{Q}$ such that $e \: \sigma \in [0,
    1]$ for all program states $\sigma : \Sigma$, command
    $\choice{e}{c_1}{c_2}$ 
    executes command $c_1$ with probability $e \: \sigma$, or $c_2$ with
    probability $1 - e \: \sigma$.
  \item \textbf{Conditioning:} Given predicate $P
    : \Sigma \rightarrow \mathbb{B}$ on program states, command
    $\obs{P}{}$ conditions the posterior distribution of the program
    on $P$.
\end{enumerate}

Additionally, command $\uniform{e}{k}$ uniformly samples a natural
number $0 \le n < e \: \sigma$ (where $\sigma$ is the current program
state) and continues execution with command $k \: n$.

\subsection{Conditional Weakest Pre-Expectation Semantics}
\label{subsec:cwp}
We follow~\cite{olmedo2018conditioning} in interpreting $\cpGCL$
programs using conditional weakest pre-expectation ($\cwp$) semantics,
a quantitative generalization of weakest precondition
semantics~\cite{dijkstra1975guarded}.  Samplers produced by \zar are
proved correct wrt.~the $\cwp$ semantics of source programs.

An \textit{expectation} is a function $f : \expectation$ mapping
program states to the nonnegative reals extended with $+\infty$. The
$\cwp$ semantics interprets programs as expectation transformers:
Given a \textit{post-expectation} $f : \expectation$ and program $c
: \cpGCL$, the weakest pre-expectation $\mathsf{wp} \: c \: f
: \expectation$ is a function mapping program states $\sigma : \Sigma$
to the expected value of $f$ over terminal states of $c$ given initial
state $\sigma$.

Given a predicate $Q : \Sigma \rightarrow \bool$, the weakest
pre-expectation of a program $c$ wrt.~post-expectation $[Q]$ maps
states $\sigma : \Sigma$ to the probability that $c$, when executed
from initial state $\sigma$, terminates in a final state satisfying
$Q$. Thus, $\cwp$ can be used to answer probabilistic queries about
the execution behavior of programs. For more background on weakest
pre-expectation semantics, see~\cite[Chapters
2-4]{kaminski2019advanced}.

The $\cwp$ semantics is defined in terms of the more primitive weakest
pre-expectation ($\wp$) and weakest liberal pre-expectation ($\wlp$)
expectation transformers. Definitions~\ref{def:wp} and~\ref{def:wlp}
give generalized variants of $\wp$ and $\wlp$ respectively, extended
with an additional Boolean parameter
controlling how observation failure is handled (where $\mathbf{0}$ and
$\mathbf{1}$ denote the constant expectations $\lambda \_.\:0$ and
$\lambda \_.\:1$, respectively, and $F^n$ denotes the $n$-fold
composition of functional $F$).

\begin{definition}[\link{cwp}{\cwpwpz}{$\mathsf{wp_b}$}]
  \label{def:wp}

  For $\mathsf{b} : \mathbb{B}$, $c : \cpGCL$, $f
  : \Sigma \rightarrow \eR$, and $\sigma : \Sigma$, define
  $\mathsf{wp_b} \: c \: f \: \sigma : \eR$ by induction on $c$:
  \begin{center}
    \setlength{\tabcolsep}{3pt}
    \begin{tabular}{l l l l}
      \rowcolor{lightgray}
      \multicolumn{4}{l}{$\link{cwp}{\cwpwpz}{\mathsf{wp_b}}
      : \cpGCL \rightarrow
      (\expectation) \rightarrow \Sigma \rightarrow \eR$} \\
      \hline
      $\cons{skip}{}$ & $f$ & $\triangleq$ & $f$ \\
      $x \leftarrow e$ & $f$ & $\triangleq$ & $f$[$x$/$e$] \\
      $\cons{observe}{} \: e$ & $f$ & $\triangleq$ & $[e] \cdot f +
      [\neg e \land \mathsf{b}]$ \\
      $c_1 ; \: c_2$ & $f$ & $\triangleq$ & $\mathsf{wp_b} \: c_1 \:
      (\mathsf{wp_b} \: c_2 \: f)$ \\
      $\cons{if}{} \: e \: \cons{then}{} \: c_1 \: \cons{else}{} \:
      c_2$ & $f$ & $\triangleq$ & $[e] \cdot \mathsf{wp_b} \: c_1 \: f
      + [\lnot e] \cdot \mathsf{wp_b} \: c_2 \: f$ \\
      $\{ \: c_1 \: \} \: [p] \: \{ \: c_2 \: \}$ & $f$ & $\triangleq$
      & $p \cdot \mathsf{wp_b} \: c_1 \: f + (\mathbf{1} -
      p) \cdot \mathsf{wp_b} \: c_2 \: f$ \\
      $\cons{uniform}{} \: e \: k$ & $f$ & $\triangleq$ &
      $\lambda \sigma. \: \frac{1}{e \: \sigma} \sum_{i=0}^{e \: \sigma
      - 1}{\mathsf{wp_b} \: (k \: i) \: f \: \sigma}$ \\
      $\cons{while}{} \: e \: \cons{do}{} \: c \: \cons{end}{}$ & $f$
      & $\triangleq$ & $\sup{(F^n \: \mathbf{0})}$, where \\
      & & & $F \: g \triangleq [e] \cdot \mathsf{wp_b} \: c \: g +
      [\lnot e] \cdot f$
    \end{tabular}
  \end{center}
\end{definition}

\begin{definition}[\link{cwp}{\cwpwlpz}{$\mathsf{wlp_b}$}]
  \label{def:wlp}

  For $b : \mathbb{B}$, $c : \cpGCL$, $f : \Sigma \rightarrow \beR$ (a
  bounded expectation), and $\sigma : \Sigma$, define
  $\mathsf{wlp_b} \: c \: f \: \sigma : \beR$ inductively on
  $c$ (we list only the $\cons{while}{}$ case as the rest are like
  $\wp$, \textit{mutatis mutandis}):
  \begin{center}
    \setlength{\tabcolsep}{3pt}
    \begin{tabular}{l l l l}
      \rowcolor{lightgray}
      \multicolumn{4}{l}{$\link{cwp}{\cwpwlpz}{\mathsf{wlp_b}}
      : \cpGCL \rightarrow
      (\bexpectation) \rightarrow \Sigma \rightarrow \beR$} \\
      \hline
      $\cons{while}{} \: e \: \cons{do}{} \: c \: \cons{end}{}$ & $f$
      & $\triangleq$ & $\inf{(F^n \: \mathbf{1})}$, where \\
      & & & $F \: g \triangleq [e] \cdot \mathsf{wlp_b} \: c \: g +
      [\lnot e] \cdot f$
    \end{tabular}
  \end{center}
\end{definition}

We often omit the subscript when $b=\false$ as $\mathsf{wp_{\false}}$
($\mathsf{wlp_{\false}}$) coincides with the classic definition of
$\wp$ ($\wlp$). The $\mathsf{sup}$ ($\mathsf{inf}$) operation is
defined wrt.~the pointwise lifting to expectations of the standard
order on $\eR$ (i.e., $f \sqsubseteq g \iff \forall \sigma,
f \: \sigma \le g \: \sigma$ for expectations $f$ and $g$). The
parameter $\mathsf{b}$ controls whether or not to include the
probability mass of observation failure, so that we have
$\mathsf{wp_b} \: c \: f + \mathsf{wlp_{\neg b}} \: c \: (\mathbf{1} -
f) = \mathbf{1} \text{, where } f \sqsubseteq \mathbf{1}$
(the \textit{invariant sum} property). $\wp$ and $\wlp$ differ as
follows:

\smallskip
$\bullet$\ $\wp$ encodes \textit{total} program correctness.
  When posing a query over predicate $Q$ using $\wp$, we are asking
  ``what is the probability that the program terminates \textit{and}
  does so in a state satisfying $Q$?''. Divergent execution paths
  (those which never terminate) contribute nothing to the
  pre-expectation.

\smallskip
$\bullet$\ $\wlp$ encodes \textit{partial} program
  correctness. When posing a query over predicate $Q$ using $\wlp$, we
  are asking ``what is the probability that the program either
  diverges \textit{or} terminates in a state satisfying
  $Q$?''. Divergent paths contribute their full probability mass to
  the weakest liberal pre-expectation.

\smallskip
Furthermore, $\wlp$ is defined only on \textit{bounded expectations}
$f : \Sigma \rightarrow \beR$ as it is only meaningful for
probabilities. It follows that $\wp$ and $\wlp$ coincide for bounded
expectations on terminating programs (whether absolutely or almost
surely).

The conditional weakest pre-expectation of expectation $f
: \expectation$ with respect to program $c : \cpGCL$ is then defined
following the approach of~\cite{olmedo2018conditioning}:
\begin{definition}[\link{cwp}{\cwpcwp}{$\mathsf{cwp}$}]
  \label{def:cwp}

  For $c : \cpGCL$ a command and $f : \Sigma \rightarrow \eR$ an expectation,
  $\mathsf{cwp} \: c \: f
  : \expectation \triangleq \frac{\mathsf{wp_{\false}} \: c \:
  f}{\mathsf{wlp_{\false}} \: c \: \mathbf{1}}$.
\end{definition}

\section{Compiling $\cpGCL$}
\label{sec:compiling-cpGCL}

We compile $\cpGCL$ commands $c$ to samplers that we prove correct
(cf.~Section~\ref{sec:correctness-of-sampling}) wrt.~the $\cwp$
semantics of $c$. First, we give an intermediate representation
(Sections~\ref{subsec:cf-trees},~\ref{subsec:cf-tree-semantics})---\textit{choice-fix}
(CF) tree---to which a command $c$ is compiled
(Section~\ref{subsec:compiling-to-cf-trees}). Next, the generated CF
tree is ``unfolded'' into a coinductive \textit{interaction tree}
(Section~\ref{subsec:generating-itrees}) implementing a sampler from
the posterior distribution denoted by $c$. Compiler correctness is
established by the semantics preservation theorem
(Theorem~\ref{theorem:cf-tree-compiler-correctness}).

\subsection{Choice-Fix Trees}
\label{subsec:cf-trees}

Choice-fix (CF) trees are named for their two nonleaf constructors:
$\cons{Choice}{}$ nodes for probabilistic choice and $\cons{Fix}{}$
nodes for encoding loops.

\begin{definition}[\link{tree}{\cwptree}{CF trees}]
  \label{def:cf-trees}

  Define the type of CF trees $\cftree$ inductively as:
  \begin{mathpar}
    \inferrule [\link{tree}{\cwpLeaf}{cf-leaf}]
               { \sigma : \Sigma }
               { \cons{Leaf}{} \: \sigma : \cftree }
    \and
    \inferrule [\link{tree}{\cwpFail}{cf-fail}]
               { }
               { \cons{Fail}{} : \cftree }
    \and
    \inferrule [\link{tree}{\cwpChoice}{cf-choice}]
               { p : \mathbb{Q} \\
                 0 \le p \le 1 \\
                 k : \bool \rightarrow \cftree }
               { \cons{Choice}{} \: p \: k : \cftree }
    \and
    \inferrule [\link{tree}{\cwpFix}{cf-fix}]
               { \sigma : \Sigma \\
                 e : \Sigma \rightarrow \bool \\
                 g : \Sigma \rightarrow \cftree \\
                 k : \Sigma \rightarrow \cftree }
               { \cons{Fix}{} \: \sigma \: e \: g \: k : \cftree }
  \end{mathpar}
\end{definition}

\begin{wrapfigure}{r}{0.62\columnwidth}

  \begin{lstlisting}[language=cpGCL,mathescape=true]
$\cons{Choice}{} \: p \: (\lambda b_0.$
    if $b_0$ then
      $\cons{Fix}{} \: \{ h \mapsto 0, b \mapsto \true \} \: (\lambda \sigma. \: \sigma[b])$
        $(\lambda \sigma. \: \cons{Choice}{} \: p \: (\lambda b'.$ if $b'$
           then $\cons{Leaf}{} \: (\sigma[h \mapsto h + 1, b \mapsto \true])$
           else $\cons{Leaf}{} \: (\sigma[h \mapsto h + 1, b \mapsto \false])))$
        $(\lambda \sigma.$ if $\sigma \: h$ is prime then $\cons{Leaf}{} \: \sigma$ else $\cons{Fail}{})$
    else $\cons{Fail}{})$
  \end{lstlisting}

  \caption{CF tree term representation of
  Prog.~\ref{prog:intro-primes}.}

  \label{fig:primes-cf-tree}
\end{wrapfigure}

$\cons{Leaf}{} \: \sigma$ denotes the end of a program execution with
terminal state $\sigma$. $\cons{Fail}{}$ denotes a program execution
in which an observed predicate (via the $\cons{observe}{}$ command) is
violated. $\cons{Choice}{} \: p \: k$ represents a probabilistic
binary choice between two subtrees where rational bias $p \in [0, 1]$
denotes the probability of ``heads'' or ``going left'' (and $1 - p$
the probability of ``tails'' or ``going right'').

Lastly, $\cons{Fix}{} \: \sigma$ $e$ $g$ $k$ encodes a loop with
initial state $\sigma$, guard condition $e$, body generator $g$, and
continuation $k$, and should be understood operationally as follows:
Starting with initial CF tree $\cons{Leaf}{} \: \sigma$, repeatedly
extend the leaves of the tree constructed thus far via either the
generating function $g$ (when $e \: \sigma = \true$) or continuation
$k$ (when $e \: \sigma = \false$). That is, $\sigma$ is the initial
state of the loop, $e$ is the guard condition of the loop, $g$ is the
generating function of the body of the loop, and $k$ is the
continuation of the program after exiting the
loop. Figure~\ref{fig:primes-cf-tree} shows the CF tree representation
of the `$\mathsf{primes}$' program from
Figure~\ref{prog:intro-primes}.

\subsection{CF Tree Semantics}
\label{subsec:cf-tree-semantics}

The \textit{inference} (or \textit{twp}) semantics of CF trees is
defined analogously to the $\mathsf{cwp}$ semantics of $\cpGCL$.  The
expression $\mathsf{twp_\false} \: t \: f$ denotes the expected value
of expectation $f$ over CF tree $t$. When $\mathsf{b}=\true$, $\twp$
additionally includes the probability mass of observation failure (the
$\cons{Fail}{}$ case of Definition~\ref{def:twp-b}).

\begin{definition}[\link{tcwp}{\cwptwpz}{$\mathsf{twp_b}$}]
  \label{def:twp-b}
  
  For $t : \cftree$ a CF tree and $f : \expectation$ an
  expectation, define $\mathsf{twp_b} \: t \: f : \eR$
  by induction on $t$:
  \begin{center}
    \setlength{\tabcolsep}{3pt}
    \begin{tabular}{l l l l}
      \rowcolor{lightgray}
      \multicolumn{4}{l}{$\link{tcwp}{\cwptwpz}{\mathsf{twp_b}}
      : \cftree \rightarrow (\expectation) \rightarrow \eR$} \\    
      \hline
      $\cons{Leaf}{} \: \sigma$ & $f$ & $\triangleq$ &
      $f \: \sigma$ \\
      $\cons{Fail}{}$ & \_ & $\triangleq$ & [$\mathsf{b}$] \\
      $\cons{Choice}{} \: p \: k$ & $f$ & $\triangleq$ &
      $p \cdot \mathsf{twp_b} \: (k \: \true) \: f + (1 -
      p) \cdot \mathsf{twp_b} \: (k \: \false) \: f$ \\
      $\cons{Fix}{} \: \sigma_0 \: e \: g \: k$ & $f$ & $\triangleq$ &
      $\sup{(F^n \: \mathbf{0})} \: \sigma_0$, where \\
      & & & $F \: h \: \sigma \triangleq \text{if } e \: \sigma \text{
      then } \mathsf{twp_b} \: (g \: \sigma) \: h \text{ else
      } \mathsf{twp_b} \: (k \: \sigma) \: f$
    \end{tabular}
  \end{center}

\end{definition}

The ``liberal'' variant $\mathsf{twlp_b} \: t \: f$ of inference
semantics denotes the expected value of expectation $f$ over CF tree
$t$ plus the probability mass of divergence (and plus the mass of
observation failure when $b = \true$). Only the $\cons{Fix}{}$ case is
shown here.

\begin{definition}[\link{tcwp}{\cwptwlpz}{$\mathsf{twlp_b}$}]
  \label{def:twlp-b}
  
  For $t : \cftree$ a CF tree and $f : \bexpectation$ a bounded
  expectation, define $\mathsf{twlp_b} \: t \: f : \beR$ inductively on $t$ as:
  \begin{center}
    \setlength{\tabcolsep}{3pt}
    \begin{tabular}{l l l l}
      \rowcolor{lightgray}
      \multicolumn{4}{l}{$\link{tcwp}{\cwptwlpz}{\mathsf{twlp_b}}
      : \cftree \rightarrow (\bexpectation) \rightarrow \beR$} \\
      \hline
      $\cons{Fix}{} \: \sigma_0 \: e \: g \: k$ & $f$ & $\triangleq$ &
      $\inf{(F^n \: \mathbf{1})} \: \sigma_0$, where \\
      & & & $F \: h \: \sigma \triangleq \text{if } e \: \sigma \text{
      then } \mathsf{twlp_b} \: (g \: \sigma) \: h \text{ else
      } \mathsf{twlp_b} \: (k \: \sigma) \: f$
    \end{tabular}
  \end{center}
\end{definition}

The conditional ($\mathsf{tcwp}$) semantics for CF trees then matches
$\cwp$ (Def.~\ref{def:cwp}):
  
\begin{definition}[\link{tcwp}{\cwptcwp}{$\mathsf{tcwp}$}]
  \label{def:tcwp}

  For $t : \cftree$ a CF tree and $f : \expectation$ an expectation,
  $\mathsf{tcwp} \: t \: f
  : \eR \triangleq \frac{\mathsf{twp_\false} \: t \:
  f}{\mathsf{twlp_\false} \: t \: \mathbf{1}}$.
\end{definition}

Our intent is that the tcwp semantics of the CF tree representation of
a cpGCL program $c$ should coincide exactly with the cwp semantics of
$c$. The following section presents a semantics-preserving compiler
from cpGCL to CF trees.

\subsection{Compiling to CF Trees}
\label{subsec:compiling-to-cf-trees}

A command $c : \cpGCL$ is compiled to a function $\llbracket
c \rrbracket : \Sigma \rightarrow \cftree$ mapping initial state
$\sigma : \Sigma$ to the CF tree encoding the sampling semantics of
$c$ starting from $\sigma$. The operator `$\bind{}$' used for
compiling sequenced commands denotes \link{tree}{\cwptreezbind}{bind}
in the CF tree monad.

\begin{definition}[\link{compile}{\cwpcompile}{$\llbracket \cdot \rrbracket$}]
  \label{def:compile}

  For $c : \cpGCL$ a command and $\sigma : \Sigma$ a program state,
  define $\llbracket c \rrbracket \: \sigma : \cftree$ inductively on $c$ by:
  \begin{center}
    \setlength{\tabcolsep}{3pt}
    \begin{tabular}{l l l l}
      \rowcolor{lightgray}
      \multicolumn{4}{l}{$\link{compile}{\cwpcompile}{\llbracket \cdot \rrbracket}
      : \cpGCL \rightarrow \Sigma \rightarrow \cftree$} \\
      \hline

      $\cons{skip}{}$ & $\sigma$ & $\triangleq$ &
      $\cons{Leaf}{} \: \sigma$ \\

      $\assign{x}{e}$ & $\sigma$ & $\triangleq$ &
      $\cons{Leaf}{} \: \sigma[x \mapsto e \: \sigma]$ \\

      $\obs{e}$ & $\sigma$ & $\triangleq$ & $\text{if }
      e \: \sigma \text{ then } \cons{Leaf \: \sigma}{} \text{ else
      } \cons{Fail}{}$ \\

      $\seq{c_1}{c_2}$ & $\sigma$ & $\triangleq$ & $\llbracket
      c_1 \rrbracket \: \sigma \bind{} \llbracket c_2 \rrbracket$ \\

      $\ite{e}{c_1}{c_2}$ & $\sigma$ & $\triangleq$ & $\text{if }
      e \: \sigma \text{ then } \llbracket
      c_1 \rrbracket \: \sigma \text{ else } \llbracket
      c_2 \rrbracket \: \sigma$ \\

      $\choice{e}{c_1}{c_2}$ & $\sigma$ & $\triangleq$ &
      $\cons{Choice}{} \: (e \: \sigma) \: (\lambda b. \: \text{if }
      b \text{ then } \llbracket c1 \rrbracket \: \sigma \text{ else
      } \llbracket c2 \rrbracket \: \sigma)$ \\

      $\uniform{e}{k}$ & $\sigma$ & $\triangleq$ &
      $\mathsf{uniform\_tree} \: (e \: \sigma) \bind{} \lambda
      n. \: \llbracket k \: n \rrbracket \: \sigma$ \\

      $\while{e}{c}$ & $\sigma$ & $\triangleq$ &
      $\cons{Fix}{} \: \sigma \: e \: \llbracket
      c \rrbracket \: \cons{Leaf}{}$ \\

    \end{tabular}
  \end{center}
\end{definition}

$\mathsf{uniform\_tree}$ is a generic construction for uniform
distributions over a fixed range of natural numbers, the specification
of which is captured by the following lemma:

\begin{lemma}[\link{uniform}{\cwptwpzuniformztreezsum}{Uniform tree correctness}]
  \label{lemma:uniform-tree}

  Let $0 < n : \nat$ and $f : \nat \rightarrow \eR$ an $\eR$-valued
  function on $\nat$. Then, $\mathsf{twp_\false} \:
  (\mathsf{uniform\_tree} \: n) \: f
  = \frac{1}{n} \sum_{i=0}^{n-1}{f \: i}. \qed$
\end{lemma}

By setting $f = [\lambda m. \: m = k]$, we obtain as an immediate
consequence of Lemma~\ref{lemma:uniform-tree} that
$\mathsf{twp_\false} \: (\mathsf{uniform\_tree} \: n) \: [\lambda
m. \: m = k] = \frac{1}{n}$ for all $k < n$, or in other words, that
$\mathsf{uniform\_tree}$ is truly uniformly distributed. The
compiler phase is then proved correct by the following theorem establishing
the correspondence of the $\cwp$ semantics of $\cpGCL$ programs with
the $\tcwp$ semantics of the CF trees generated from them.

\begin{theorem}[\link{cwp\_tcwp}{\cwpcwpztcwp}{CF tree compiler correctness}]
  \label{theorem:cf-tree-compiler-correctness}

  Let $c : \cpGCL$, $f : \expectation$, and $\sigma : \Sigma$. Then,
  $\tcwp \: (\llbracket c \rrbracket \: \sigma) \: f = \cwp \: c \:
  f \: \sigma. \qed$
\end{theorem}

\paragraph*{Debiasing CF trees}
CF trees generated by Def.~\ref{def:compile} may have arbitrary $p \in
[0,1]$ bias values at choice nodes. To move toward the (uniform)
random bit model, we apply a bias-elimination transformation
$\link{debias}{\cwpdebias}{\mathsf{debias}
: \cftree \rightarrow \cftree}$ that uses the $\mathsf{uniform\_tree}$
construction described above to replace probabilistic choices by
semantically equivalent fair coin-flipping schemes. The resulting CF
trees have $p=\frac{1}{2}$ at all choice nodes (we say that such trees
are \textit{unbiased}). Figure~\ref{fig:debias} shows how a choice
node with bias $\frac{2}{3}$ is reduced to an equivalent unbiased CF
tree. A more detailed explanation of the $\mathsf{debias}$ algorithm
is given in Appendix~\ref{app:debias}. The essential results for
$\mathsf{debias}$ are: $\mathsf{debias}$ preserves tcwp semantics, and
produces CF trees in which all choice nodes are unbiased.

\begin{figure}[t]
  \centering
  \begin{subfigure}[t]{.5\textwidth}\centering
    \scalebox{0.8}{\input{biased2.tikz}}
    \caption{Biased choice with Pr($\true$) $= \frac{2}{3}$}
    \label{fig:two-thirds-choice}
  \end{subfigure}%
  \begin{subfigure}[t]{.5\textwidth}\centering
    \scalebox{0.8}{\input{debiased2.tikz}}
    \caption{Debiased tree with Pr($\true$) $= \frac{2}{3}$}
    \label{fig:two-thirds-debiased}
  \end{subfigure}

  \caption{$\cons{Choice}{}$ CF tree with $Pr(\true) = \frac{2}{3}$
  (left) and corresponding debiased CF tree (right).}

  \label{fig:debias}
\end{figure}
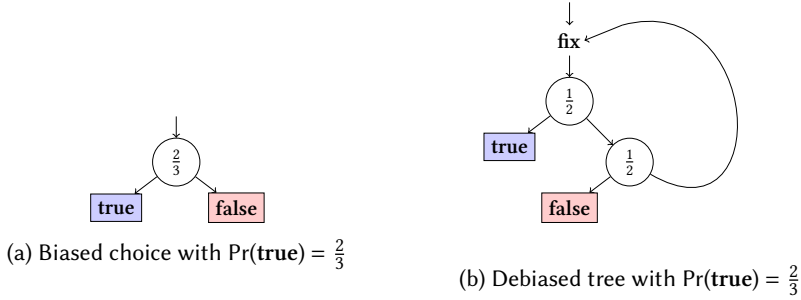

\begin{theorem}[\link{debias}{\cwptcwpzdebias}{$\mathsf{debias}$ is sound}]
  \label{theorem:debias-sound}

  Let $t : \cftree$ be a CF tree and $f : \expectation$ an
  expectation. Then, $\tcwp \: (\mathsf{debias} \: t) \: f = \tcwp \:
  t \: f. \qed$
\end{theorem}

\begin{theorem}[\link{debias}{\cwptreezunbiasedzdebias}{$\mathsf{debias}$ produces unbiased trees}]
  \label{theorem:debias-unbiased}

  Let $t : \cftree$ be a CF tree. Then, $p = \frac{1}{2}$ for every
  $\: \cons{Choice}{} \: p \: k$ in $\mathsf{debias} \: t$. $\qed$
\end{theorem}

\paragraph*{The Need for Coinduction}
Debiased CF trees are close to being executable samplers in the random
bit model. However, since they permit the existence of infinite
execution paths (e.g., when $b=\true$ ad infinitum in
Program~\ref{prog:intro-primes}), and hence denote sampling processes
that can't be expected in general to terminate absolutely, we must
first pass from the \textit{inductive} CF tree encoding of samplers to
an infinitary \textit{coinductive} encoding.

\subsection{Generating Interaction Trees}
\label{subsec:generating-itrees}

Interaction trees~\cite{xia2019interaction} (ITrees) are a
general-purpose coinductive data structure for modeling effectful
(co-)recursive programs that interact with their
environments. The \href{https://github.com/DeepSpec/InteractionTrees}{$\mathsf{coq}$-$\mathsf{itree}$}
library provides a suite of combinators for constructing ITrees along
with a collection of formal principles for reasoning about their
equivalence. An interaction tree computation performs an effect by
raising an event (which may carry data) that is then handled by the
environment, possibly providing data in return. This section shows how
to generate executable interaction tree samplers from CF trees.

\paragraph*{Interaction Tree Syntax}
Interaction trees are parameterized by an event functor $E
: \mathsf{Type} \rightarrow \mathsf{Type}$ that specifies the kinds of
interactions an ITree process can have with its environment. In our
case (Definition~\ref{def:itrees}), there is only one kind of
interaction: the sampler may query the environment for a single
randomly generated bit. Thus the event functor $\mathsf{boolE}$ has a
single constructor $\cons{GetBool}{}$ taking zero arguments, with type
index $\bool$ indicating that the environment's response should be a
single bit.

\begin{definition}[\link{itree}{\cwpitreezsampler}{$\itree{A}$}]
  \label{def:itrees}

  Define $\itree{A}$ -- the type of interaction trees with event functor
  $\mathsf{boolE}$ and element type $A$ -- coinductively by:
  \begin{mathpar}
    \inferrule [\link{itree}{\cwpGetBool}{boolE-GetBool}]
               {  }
               { \cons{GetBool}{} : \mathsf{boolE} \: \bool }
    \and
    \inferrule [itree-ret]
               { a : A }
               { \cons{Ret}{} \: a : \itree{A} }
    \and
    \inferrule [itree-tau]
               { t : \itree{A} }
               { \cons{Tau}{} \: t : \itree{A} }
    \and
    \inferrule [itree-vis]
               { k : \bool \rightarrow \itree{A} }
               { \cons{Vis}{} \: \cons{GetBool}{} \: k : \itree{A} }
  \end{mathpar}
\end{definition}

Unfolding a CF tree to an interaction tree proceeds in two steps:
\begin{enumerate}
  \item Generating an ITree $t : \itree{(\unit + \Sigma)}$ by
  induction on the input CF tree, and then
  \item ``tying the knot'' on $t$ to produce the final ITree of type
    $\itree{\Sigma}$.
\end{enumerate}

The LHS of the sum type $\unit + \Sigma$ is used to encode observation
failure. Since ITrees have just one kind of terminal constructor
($\cons{Ret}{}$) to CF trees' two ($\cons{Leaf}{}$ and
$\cons{Fail}{}$), we translate $\cons{Fail}{}$ nodes to
$\cons{Ret}{} \: (\mathsf{inl} \: \tt)$, and nodes of the form
$\cons{Leaf}{} \: x$ to $\cons{Ret}{} \: (\mathsf{inr} \: x)$, where
$\mathsf{inl}$ and $\mathsf{inr}$ are the left and right sum
injections.  $\cons{Fix}{}$ nodes are translated via application of
the $\mathsf{ITree.iter}$ combinator (see~\cite[Section
4]{xia2019interaction} on iteration with ITrees).

The first step is implemented by the function
$\mathsf{to\_itree\_open}$ (see Figure~\ref{fig:prime-itree-open}):

\begin{definition}[\link{itree}{\cwptozitreezopen}{$\mathsf{to\_itree\_open}$}]
  \label{def:to-itree-open}

  For unbiased CF tree $t : \cftree$, define
  $\mathsf{to\_itree\_open} \; t : \itree{\unit + \Sigma}$ by
  induction on $t$ as:
  \begin{center}
    \setlength{\tabcolsep}{3pt}
    \begin{tabular}{l l l}
      \rowcolor{lightgray}
      
      \multicolumn{3}{l}{$\link{itree}{\cwptozitreezopen}{\mathsf{to\_itree\_open}} \:
      : \cftree \rightarrow \itree{\unit + \Sigma}$} \\
      
      \hline
      
      $\cons{Leaf}{} \: \sigma$ & $\triangleq$ & $\cons{Ret}{} \:
      (\mathsf{inr \: \sigma})$ \\
      
      $\cons{Fail}{}$ & $\triangleq$ & $\cons{Ret}{} \:
      (\mathsf{inl} \: \tt)$ \\
      
      $\cons{Choice}{} \: \_ \: k$ & $\triangleq$ &
      $\cons{Vis}{} \: \cons{GetBool}{} \:
      (\mathsf{to\_itree\_open} \circ k)$ \\
      
      $\cons{Fix}{} \: \sigma_0 \: e \: g \: k$ & $\triangleq$ &
      $\mathsf{ITree.iter} \: (\lambda \sigma. \: \text{if }
      e \: \sigma$ then $y \leftarrow \mathsf{to\_itree\_open} \:
      (g \: \sigma) \: ;;$ \\

      & & \hspace{101pt} $\text{match } y$ with \\

      & & \hspace{101pt}
      $\mid \mathsf{inl} \: \tt \Rightarrow \cons{Ret}{} \:
      (\mathsf{inr} \: (\mathsf{inl} \: \tt))$ \\

      & & \hspace{101pt}
      $\mid \mathsf{inr} \: \sigma' \Rightarrow \cons{Ret}{} \:
      (\mathsf{inl} \: \sigma')$ \\
      
      & & \hspace{101pt} end \\

      & & \hspace{83pt} else $\mathsf{ITree.map} \: \mathsf{inr} \:
      (\mathsf{to\_itree\_open} \: (k \: \sigma))) \: \sigma_0$ \\
    \end{tabular}
  \end{center}
\end{definition}

ITrees produced by $\mathsf{to\_itree\_open}$ treat observation
failure as a terminal state with unit value $\tt$
(Figure~\ref{fig:prime-itree-open}). The following definition
$\mathsf{tie\_itree}$ corecursively ``ties the
knot''~\cite{elkins2021knot} through the left side of $\unit + \Sigma$
via $\mathsf{ITree.iter}$ to produce an ITree rejection sampler that
restarts from the beginning upon observation failure
(Figure~\ref{fig:prime-itree-closed}).

\begin{definition}[\link{itree}{\cwptiezitree}{$\mathsf{tie\_itree}$}]
  For $t : \itree{\unit + \Sigma}$, define $\mathsf{tie\_itree} \: t : \itree{\Sigma}$ as:
  \[ \mathsf{tie\_itree} \: t \triangleq \mathsf{ITree.iter} \:
  (\lambda \_. \: t) \: \tt \]
\end{definition}

\begin{figure}[H]
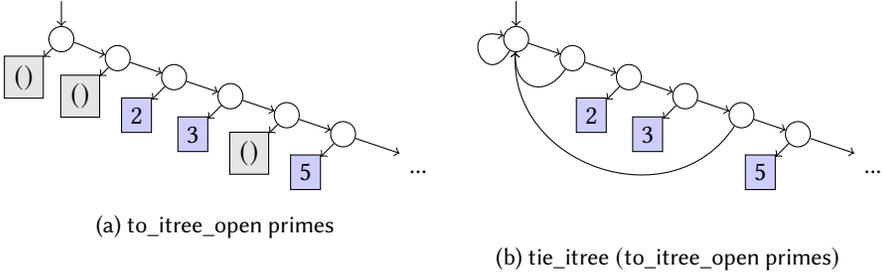

  \begin{subfigure}[t]{.5\textwidth}
    \centering
    \ctikzfig{prime_sampler3_small_open}
    \caption{$\mathsf{to\_itree\_open} \: \mathsf{primes}$}
    \label{fig:prime-itree-open}
  \end{subfigure}%
  \begin{subfigure}[t]{.5\textwidth}
    \centering
    \ctikzfig{prime_sampler3_small}
    \caption{$\mathsf{tie\_itree} \: (\mathsf{to\_itree\_open} \: \mathsf{primes})$}
    \label{fig:prime-itree-closed}
  \end{subfigure}

  \caption{Interaction trees generated by
  `$\mathsf{to\_itree\_open} \: \mathsf{primes}$' (left) and then by
  ``tying the knot'' via `$\mathsf{tie\_itree}$' (right), where
  `$\mathsf{primes}$' is the $\cpGCL$ program in
  Figure~\ref{prog:intro-primes}.}

  \label{fig:prime-itree}
\end{figure}

\paragraph*{Interaction tree semantics}
We wish to define an analogue $\mathsf{itwp}$ of the $\cwp$ semantics
for ITree samplers and prove the correctness of ITree generation with
respect to it, such that $\mathsf{itwp} \: f \:
(\mathsf{tie\_itree} \: (\mathsf{to\_itree\_open} \: t)) = \tcwp \:
t \: f$ for all $t : \cftree$ and $f : \expectation$. This turns out,
however, to be difficult due to the lack of induction principle for
ITrees.

To overcome the problem, we note that ITree samplers form
an \link{aCPO}{\cwpaCPO}{\textit{algebraic CPO}}~\cite[Chapter
5]{gunter1992semantics}, i.e., a domain in which all elements can be
obtained as suprema of $\omega$-chains of finite
approximations. Moreover, the types $\eR$ of extended reals and
$\prop$ of propositions are CPOs. We exploit these observations to
provide a special kind of \link{aCPO}{\cwpco}{induction principle} for
coinductive trees (see ~\cite[Lemma 5.24]{gunter1992semantics}). Such
a principle enables the definition of
Scott-continuous~\cite{abramsky1994domain} functions
like \link{itree}{\cwpitwp}{$\mathsf{itwp}$}, and gives rise to a
powerful suite of proof principles for reasoning about them via
reduction to inductive proofs over inductive structures (for which Coq
is much better suited than for coinduction). While details of this
framework are outside the scope of this paper, they are implemented in
the
accompanying \href{https://github.com/bagnalla/zar/tree/release-pldi23}{Coq
sources}.

\paragraph*{End-to-end compiler}
The compiler pipeline steps are composed via
$\mathsf{cpGCL\_to\_itree}$ (with $\mathsf{elim\_choices}$ reducing
rationals and coalescing duplicate $\cons{Leaf}{}$ nodes) and proved
correct by Theorem~\ref{theorem:compiler-correctness} (with positivity
constraint on $\mathsf{wlp_{\false}} \: c \: \mathbf{1}\: \sigma$
assuring that the program doesn't condition on contradictory
observations):

\begin{definition}[\link{itree}{\cwpcpGCLztozitree}{$\mathsf{cpGCL\_to\_itree}$}]
  
  For $c : \cpGCL$ and $\sigma : \Sigma$, define
  \[ \mathsf{cpGCL\_to\_itree} \:
  c \: \sigma \triangleq \mathsf{tie\_itree} \:
  (\mathsf{to\_itree\_open} \: (\mathsf{debias} \:
  (\mathsf{elim\_choices} \: (\mathsf{compile} \: c \: \sigma)))) \]
\end{definition}

\begin{theorem}[\link{itree}{\cwpcwpzitwp}{Compiler Correctness}]
  \label{theorem:compiler-correctness}
  
  Let $c : \cpGCL$, $f : \expectation$, and $\sigma : \Sigma$ such
  that $0 < \mathsf{wlp_\false} \: c \: \mathbf{1} \: \sigma$. Then,
  $ \cwp \: c \: f \: \sigma = \mathsf{itwp} \: f \:
  (\mathsf{cpGCL\_to\_itree} \: c \: \sigma). \qed$
\end{theorem}

Theorem~\ref{theorem:compiler-correctness} establishes semantics
preservation of the compiler pipeline wrt.~$\mathsf{itwp}$, but
doesn't directly guarantee properties of samples generated by the
resulting ITrees. Drawing on basic measure
theory~\cite{halmos2013measure} and the theory of uniform-distribution
modulo 1~\cite{weyl1916gleichverteilung,kuipers2012uniform}, the next section extends the
result of Theorem~\ref{theorem:compiler-correctness} to show that
sequences of generated samples are equidistributed wrt.~the $\cwp$
semantics of their source programs.

\section{Correctness of Sampling}
\label{sec:correctness-of-sampling}

Given a suitable source of \iid randomness
(Section~\ref{subsec:source-of-randomness}), a sampler for program $c
: \cpGCL$ is correct if it produces a sequence $x_n
: \nat \rightarrow \Sigma$ such that for any observation $Q
: \Sigma \rightarrow \prop$ over terminal program states,
the \textit{proportion of samples} falling within $Q$ asymptotically
converges to the expected value of $[Q]$ (i.e., the probability of
$Q$) according to $c$'s $\cwp$ semantics.  In other words, a sampler
is correct when the samples it produces
are \textit{equidistributed}~\cite{BecherG22} wrt. $\cwp$.

This section formalizes the notion of equidistribution described above
and proves the main sampling equidistribution theorem
(Theorem~\ref{theorem:cpGCL-equidistribution},
Section~\ref{subsec:equidistribution}).  We first clarify what is
meant by ``a suitable source of randomness''
(Section~\ref{subsec:source-of-randomness}) and then re-cast the
problem of inference as that of computing a measure
(Section~\ref{subsec:inference-as-measure}).

\subsection {The Source of Randomness}
\label{subsec:source-of-randomness}

We assume access to a stream of uniformly distributed bits.
The \textit{Cantor space} of countable sequences of bits
(\textit{bitstreams}), denoted $2^\nat$, is modeled by the coinductive
type $\mathsf{Stream} \: \bool$. We interpret samplers as measurable
functions from $2^\nat$ to the sample space $\Sigma$. To do so we
first turn $2^\nat$ into a measurable space by equipping it with a
measure.

\paragraph*{Bisecting the Unit Interval}
To help visualize the measure on $2^\nat$, consider
(Figure~\ref{fig:bisection}) the bisection scheme for identifying
strings of bits (e.g., ``0'', ``01'', ``011'', etc.) with dyadic
subintervals of the unit interval $[0,1]$. Let $I(\omega)$ denote the
interval corresponding to string $\omega$, and $B(\omega) \subseteq
2^\nat$ the \textit{basic set} of bitstreams with prefix $\omega$
(i.e., $B(\omega) = \{s \mid \omega \sqsubseteq s\}$ where
`$\sqsubseteq$' is the prefix order). We arrange for
the \textit{measure} of $B(\omega)$ (denoted $\mu_\Omega(B(\omega))$)
to be equal to the \textit{length} of $I(\omega)$: exactly $2^{-n}$
where $n$ is the length of $\omega$. We then define the \textit{source
of randomness} to be the measure space $\Omega$ obtained by equipping
$2^\nat$ with measure $\mu_\Omega$ lifted to the Borel
$\sigma$-algebra $\Sigma_\Omega$ of countable unions and complements
of basic sets, coinciding with the standard Lebesgue measure $\lambda$
on the Borel $\sigma$-algebra generated by subintervals of $[0, 1]$.

\begin{figure}[t]
  \centering
  \begin{subfigure}{.5\textwidth}
  \centering
  \includegraphics[width=1\linewidth]{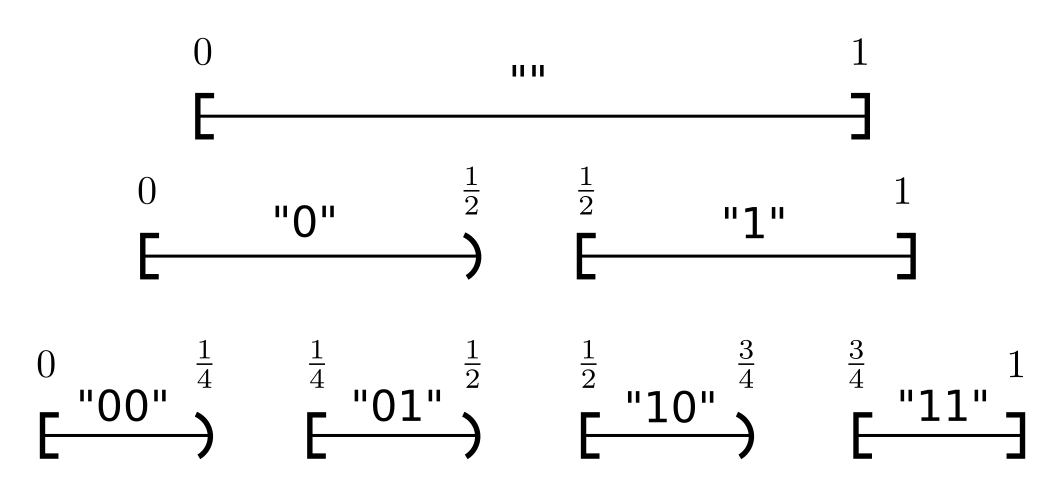}
  \caption{Bisection scheme identifying bitstrings with dyadic
    subintervals of $[0,1]$}
  \label{fig:bisection}
  \end{subfigure}%
  \begin{subfigure}{.5\textwidth}
  \begin{figure}[H]
  \centering
  \begin{subfigure}[t]{.5\textwidth}
    \centering
    \ctikzfig{two_thirds}
    \caption{ITree sampler $t_{\frac{2}{3}}$.}
    \label{fig:two-thirds-bias}
    \vspace{0.3cm}
  \end{subfigure}
  \begin{subfigure}[t]{.5\textwidth}
    \centering
    \def\svgwidth{1.2\columnwidth}
    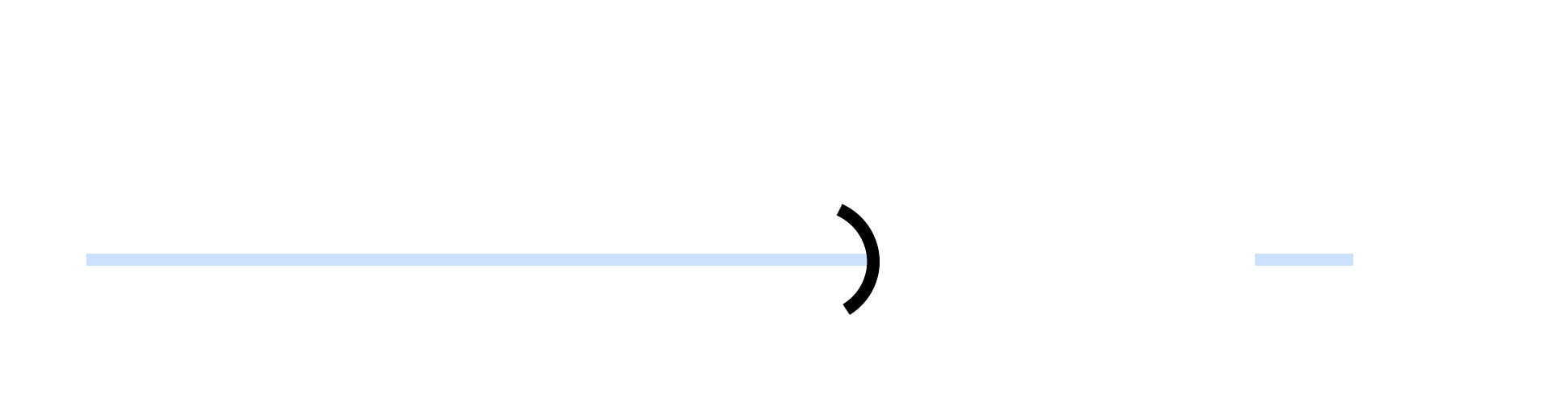
    \caption{Preimage intervals of event $\{\true\}$ under $f_{t_{\frac{2}{3}}}$.}
    \label{fig:two-thirds-bias-intervals}
  \end{subfigure}

\end{figure}
  \end{subfigure}

  \caption{Interval bisection scheme (left) and its application to
  ITree $t_{\frac{2}{3}}$ (right).}

  \label{fig:bisection-two-thirds}
\end{figure}


\subsection{Inference as Measure}
\label{subsec:inference-as-measure}

We view ITree sampler $t : \itree{\Sigma}$ as a partial measurable
function $f_t : \Omega \rightharpoonup \Sigma$ from $\Omega$ to the
sample space $(\Sigma, \Sigma_{\Sigma})$ where $\Sigma_{\Sigma}$ is
the discrete (power set) $\sigma$-algebra on the space of program
states $\Sigma$. Evaluation of $f_t$ on bitstream $s : 2^\nat$ has two
possible outcomes:
\begin{enumerate}
  \item The sampler \textit{diverges}, consuming bits ad infinitum
without ever producing a sample. In that case, we have $f_t(s)
= \bot$, i.e., $f_t$ is \textit{undefined} on $s$. We admit samplers
for which such infinite executions are permitted but occur with
probability $0$ (i.e., the set $D \subseteq 2^\nat$ of diverging
inputs has measure $0$). Or,

  \item a value $x$ is produced after consuming a finite prefix
$\omega$ of $s$ corresponding to basic set $B(\omega)$. Thus, the
function $f_t$ sends all bitstreams in $B(\omega)$ to output $x$.
\end{enumerate}

For example, consider the ITree sampler $t_{\frac{2}{3}}$ in
Figure~\ref{fig:two-thirds-bias} yielding $\true$ with probability
$\frac{2}{3}$. The \textit{preimage} set
$f_{t_{\frac{2}{3}}}^{-1}(\{\true\})$ of event $\{\true\}$ under
$f_{t_{\frac{2}{3}}}$ (i.e., the set of bitstreams sent by
$f_{t_{\frac{2}{3}}}$ to $\true$) has measure $\frac{2}{3}$. To see
why, observe that $t_{\frac{2}{3}}$ contains infinitely many disjoint
paths to $\true$ (``0'', ``100'', ``10100'', etc.), with corresponding
interval lengths $\frac{1}{2}$, $\frac{1}{8}$, $\frac{1}{32}$, etc., a
geometric series with sum
$\sum_k^\infty{\frac{1}{2} \cdot \frac{1}{4}^k} = \frac{\frac{1}{2}}{1
- \frac{1}{4}} = \frac{2}{3}$.

We can exploit this observation to let the measure of any event
$Q \subseteq \{\true, \false\}$ be equal to the measure of its
preimage under $f_{t_{\frac{2}{3}}}$, thus inducing the following
probability measure $\mu_{t_{\frac{2}{3}}}
: \{\true, \false\} \rightarrow \beR$ (where $\lambda(I)$ denotes the
length of interval $I$):

\begin{table}[H]
  \small
  \centering
  \begin{tabular}{*{7}{l}}
  
    $\mu_{t_{\frac{2}{3}}}(\emptyset)$ & $=$ &
    $\mu_\Omega(f_{t_{\frac{2}{3}}}^{-1}(\emptyset))$ & $=$ &
    $\lambda(\emptyset)$ & $=$ & $0$ \\
    
    $\mu_{t_{\frac{2}{3}}}(\{\true\})$ & $=$ &
    $\mu_\Omega(f_{t_{\frac{2}{3}}}^{-1}(\{\true\}))$ & $=$ &
    $\lambda([0, \frac{2}{3}))$ & $=$ & $\frac{2}{3}$ \\
    
    $\mu_{t_{\frac{2}{3}}}(\{\false\})$ & $=$ &
    $\mu_\Omega(f_{t_{\frac{2}{3}}}^{-1}(\{\false\}))$ & $=$ &
    $\lambda([\frac{2}{3}, 1])$ & $=$ & $\frac{1}{3}$ \\
    
    $\mu_{t_{\frac{2}{3}}}(\{\true, \false\})$ & $=$ &
    $\mu_\Omega(f_{t_{\frac{2}{3}}}^{-1}(\{\true,\false\}))$ & $=$ &
    $\lambda([0, 1])$ & $=$ & $1$ 
    
  \end{tabular}
\end{table}

\paragraph*{Computing Preimages}
We may generalize the above method to induce a probability measure
$\mu_t : \Sigma \rightarrow \beR$ from any $t : \itree{\Sigma}$ by
letting $\mu_t(Q) = \mu_\Omega(f_t^{-1}(Q))$ (the \textit{pushforward
measure} of $Q$ under $f_t$), assigning to any event $Q
: \Sigma \rightarrow \prop$ a probability equal to the measure in
$\Omega$ of its preimage under $f_t$. However, the
preimage \link{mu}{\cwpcotreezpreimage}{$f_t^{-1}(Q)$} is not easy to
determine in general, as it may be the union of infinitely many small
intervals scattered throughout $[0,1]$ in a complicated arrangement
depending on the structure of $t$. We define $f_t^{-1}(Q)$ using the
domain-theoretic machinery described in
Section~\ref{subsec:generating-itrees}.

\subsection{Equidistribution}
\label{subsec:equidistribution}

To prove correctness of samplers generated by \zar, we show that any
sequence of samples produced by them is \textit{equidistributed}
wrt.~the $\cwp$ semantics of the input programs. Our strategy is to
assume uniform distribution of the source of randomness $\Omega$, and
``push it forward'' through the sampler to obtain the desired result.
This section builds on the theory of uniform distribution modulo 1 --
in particular, the class \textit{$\Sigma_1^0$} of countable unions of
rational bounded intervals -- adapted to collections of bitstreams.

\paragraph*{Uniform distribution of $\Omega$}
We assume access to a uniformly distributed sequence of
bitstreams. But what does it mean for a sequence of bitstreams to be
uniformly distributed? We cannot simply assert that any two bitstreams
occur with equal probability because any particular bitstream occurs
with probability zero and this may be true even for nonuniform
distributions. Instead, we turn to a variation of the classic notion
of ``uniform distribution modulo 1'', generalized to the class
$\Sigma_1^0$ of subsets of $2^\nat$~\cite{kuipers2012uniform}.

A subset $U \subseteq \mathcal{P}(2^\nat)$ is said to be $\Sigma_1^0$
when it is equal to $\bigcup_i^\infty{B(\omega_i)}$ for some countable
collection $\{B(\omega_i)\}$ of basic sets.  We remark
that \link{itree}{\cwpitreezpreimage}{$f_t^{-1}(Q)$ is $\Sigma_1^0$
for all $Q : \Sigma \rightarrow \prop$ and $t : \itree{\Sigma}$}. The
required notion of uniform distribution now follows:

\begin{definition}[\link{equidistribution}{\cwpuniform}{$\Sigma_1^0$-u.d.}]
  \label{def:sigma01-ud}
  
  A sequence $\{x_i\}$ of bitstreams is \textit{$\Sigma_1^0$-uniformly
  distributed} (\textit{$\Sigma_1^0$-u.d.}) when for every $U
  : \Sigma_1^0$,
  $\lim_{n \to \infty}{\frac{1}{n} \sum_{i=0}^{n-1}{[x_i \in U]}}
  = \mu_{\Omega}(U)$.
\end{definition}

``Almost all'' sequences of bitstreams are
$\Sigma_1^0$-u.d.~\cite{BecherG22}. Moreover, $\Sigma_1^0$-u.d. has
deep connections to Martin-L\"of
randomness~\cite{martin1966definition} and Schnorr
randomness~\cite{downey2004schnorr} (cf.~\cite[Theorem
3]{BecherG22}).} We can now state the \textit{equidistribution
theorem}:

\begin{theorem}[\link{equidistribution}{\cwpcpGCLzsampleszequidistributed}{$\cpGCL$ equidistribution}]
  \label{theorem:cpGCL-equidistribution}
  
  Let $c : \cpGCL$ be a command, $\sigma : \Sigma$ a program state,
  $\{x_n\}$ a $\Sigma_1^0$-u.d. sequence of bitstreams, $Q
  : \pred{\Sigma}$ a predicate over program states, and $t
  : \itree{\Sigma} = \mathsf{cpGCL\_to\_itree \: c \: \sigma}$ the
  ITree sampler compiled from $c$. Then, the sequence $\{f_t(x_n)\}$
  of samples is $\cwp$-equidistributed wrt.~$c$:
  \[ \lim_{n \to \infty}{\frac{1}{n} \sum_{i=0}^{n-1}{[Q \: (f_t
  (x_i))]}} = \cwp \: c \: [Q] \: \sigma. \qed \]
\end{theorem}

\section{Empirical Validation}
\label{sec:empirical-validation}


This section provides empirical validation of the following
aspects of samplers compiled from $\cpGCL$ programs:

$\bullet$\ \textit{Correctness.} To validate
  Theorem~\ref{theorem:cpGCL-equidistribution}, we compare the
  empirical distribution of generated samples with the expected true
  distribution wrt.~total variation (TV) distance, Kullback-Leibler
  (KL) divergence~\cite{kullback1951information}, and
  Symmetric Mean Absolute Percentage Error (SMAPE)~\cite{armstrong1985crystal}.

$\bullet$\ \textit{Performance.}
  Although generated samplers are not
  guaranteed to be entropy-optimal (in contrast to
  OPTAS~\cite{saad2020sampling}), we measure statistics of the number
  of uniform random bits required to obtain a sample.

\smallskip
We do not verify the programs in this section wrt.~their cwp semantics
as we seek only to validate the correctness of the compilation
pipeline.

\paragraph*{OCaml Shim}

\lstset{
 language=caml,
 columns=[c]fixed,
 basicstyle=\small\ttfamily,
 keywordstyle=\bfseries,
 upquote=true,
 commentstyle=\color{OliveGreen},  
 breaklines=true,
 showstringspaces=false}

All programs in this section are compiled to verified ITree samplers
(as described in Section~\ref{subsec:generating-itrees}) and extracted
to OCaml~\cite{letouzey2008extraction, leroy:hal-00930213} for
execution by the driver code in Figure~\ref{prog:ocaml-driver}. Thus,
correctness of extracted samplers depends on the PRNG provided by the
OCaml Random module being
$\Sigma_1^0$-u.d. (Def.~\ref{def:sigma01-ud}). Sample records are
generated and written to disk for external analysis with handwritten
Python code (see,
e.g., \href{https://github.com/bagnalla/zar/tree/release-pldi23/extract/die/analyze.py}{/extract/die/analyze.py})
and statistics routines provided by
$\mathsf{scipy.stats}$~\cite{scipy2022scipystats}).

\paragraph*{Trusted Computing Base}
Our TCB includes the Coq typechecker (and therefore the OCaml compiler
and runtime), the specifications of $\cwp$ (Section~\ref{subsec:cwp})
and equidistribution (Section~\ref{subsec:equidistribution}),
and the OCaml extraction mechanism and driver code in
Figure~\ref{prog:ocaml-driver}.

\paragraph*{Empirical Evaluation}
The remainder of this section contains tables showing results of
empirical evaluation of accuracy and entropy-performance of a
collection of illustrative $\cpGCL$ programs (all wrt.~a sample size
of 100,000). In each table, the first column denotes the values taken
by the parameter of the distribution (e.g., the bias parameter $p$ for
Bernoulli, range $n$ for uniform, etc.). $\mu_x$ and $\sigma_x$ denote
the mean and standard deviation of the posterior over variable
$x$. TV, KL, and SMAPE denote the total variation distance,
Kullback-Leibler (KL) divergence~\cite{kullback1951information}, and
symmetric mean absolute percentage error~\cite{armstrong1985crystal},
respectively, of the empirical distribution wrt.~the true
distribution. $\mu_{bit}$ and $\sigma_{bit}$ denote the mean and
standard deviation of the number of uniform random bits required to
obtain a sample.

\paragraph*{Entropy Usage}
The Shannon entropy~\cite{shannon1948mathematical} of a probability
distribution provides a lower bound on the average number of uniformly
random bits required to obtain a single \iid sample. \cite{knuth1976complexity} show that an ``entropy-optimal''
sampler in the random bit model consumes no more than $2$ bits on
average than the entropy of the encoded distribution. Our samplers are
not guaranteed entropy optimal (a direction for future work).

\begin{figure}[t]
  \centering
  \begin{tabular}{c}
    \begin{lstlisting}
let _ = Random.self_init () (* Seed PRNG. *)
let rec run t =             (* t : (boolE, 'a) itree *)
  match observe t with      (* Unfold itree. *)
  | RetF x -> x             (* Produce sample. *)
  | TauF t' -> run t'       (* Skip tau node. *)
  | VisF (_, k) ->          (* Consume random bit. *)
      run (k (Obj.magic (Random.bool ())))
    \end{lstlisting}
  \end{tabular}

  \caption{OCaml shim for execution of ITree samplers. The destructor
  `$\mathsf{observe}$' (not to be confused with the $\cpGCL$ command
  of the same name) unfolds the structure of the ITree $t$.}

  \label{prog:ocaml-driver}
\end{figure}

\paragraph*{Flip}
The command $\flip{x}{p} : \cpGCL$ performs a probabilistic choice
(``flips a coin'') with probability $p : \mathbb{Q}$ of $\true$ (or
``heads'') and assigns the result to variable $x$.

\begin{definition}[\link{prelude}{\cwpflip}{flip}]
  \label{def:flip}

  For $x : \mathsf{ident}$ and $p \in [0, 1] \subseteq \mathbb{Q}$,
  define $\flip{x}{p}$ as:
  \begin{center}
    $\mathsf{flip} \: x \:
    p \triangleq \choice{p}{\assign{x}{\true}}{\assign{x}{\false}}$
  \end{center}
\end{definition}

\subsection{Dueling Coins}
\label{subsec:dueling-coins}

Figure~\ref{prog:dueling-coins} illustrates an \iid loop (unbounded
but with no loop-carried dependence) simulating a fair coin using a
biased one. The posterior distribution over $a$ is
Bernoulli($\frac{1}{2}$) for any input bias $p \in (0, 1)$. The
dueling coins illustrate a situation in which the average number of
bits required to obtain a sample ($\mu_{bit} \sim 12$ when $p
= \frac{2}{3}$ and $\mu_{bit} \sim 135$ when $p = \frac{1}{20}$, see
Table~\ref{table:dueling-coins}) substantially exceeds the entropy
(exactly $1$) of the posterior.

\begin{figure}[t]
  \centering
  \begin{subfigure}{.5\textwidth}
    \centering
    \begin{tabular}{c}
      \begin{lstlisting}[language=cpGCL,mathescape=true]
duel (p : $\mathbb{Q}$) :=
  a $\leftarrow$ $\false$; b $\leftarrow$ $\false$;
  while a = b do flip a p; flip b p; end
      \end{lstlisting}
    \end{tabular}

    \caption{\link{dueling\_coins}{\cwpduelingzcoins}{Dueling coins}
    program with bias $p \in (0, 1)$.}

    \label{prog:dueling-coins}
  \end{subfigure}\hfill%
  \begin{subfigure}{.5\textwidth}
    \centering
    \begin{tabular}{c}
    \begin{lstlisting}[language=cpGCL,mathescape=true]
die ($n : \nat$) :=
  uniform $n$ ($\lambda m.$ $x$ $\leftarrow$ $m + 1$)
    \end{lstlisting}
    \end{tabular}

  \caption{Rolling an \link{die}{\cwpdie}{$n$-sided die}.}
  \label{prog:n-sided-die}
  \end{subfigure}

  \caption{Dueling coins (left) and n-sided die (right) $\cpGCL$
  programs.}

  \label{fig:duel-geometric}
\end{figure}

\begin{table}[t]
  \centering
  \setlength{\tabcolsep}{3pt}
  \begin{tabular}{c c c c c c c c}

    $p$ & $\mu_a$ & $\sigma_a$ & TV & KL & SMAPE & $\mu_{bit}$ &
    $\sigma_{bit}$ \\

    \hline

    $2/3$ & $0.50$ & $0.50$ & $\num{2.02e-3}$ & $\num{1.20e-5}$ &
    $\num{2.02e-3}$ & $12.0$ & $9.39$ \\

    $4/5$ & $0.50$ & $0.50$ & $\num{2.16e-3}$ & $\num{1.30e-5}$ &
    $\num{2.16e-3}$ & $27.59$ & $23.49$ \\

    $1/20$ & $0.50$ & $0.50$ & $\num{2.83e-3}$ & $\num{2.30e-5}$ &
    $\num{2.83e-3}$ & $134.97$ & $129.07$

  \end{tabular}

  \caption{Accuracy and entropy usage for
  Prog.~\ref{prog:dueling-coins} with $p = \frac{2}{3}$,
  $\frac{4}{5}$, and $\frac{1}{20}$. $mu_{bit}$ and $\sigma_{bit}$
  increase as $p$ is goes further from $\frac{1}{2}$ due to increasing
  Shannon entropy of Bernoulli($p$).}

  \label{table:dueling-coins}
\end{table}

\subsection{Geometric Primes}
\label{subsec:geometric-primes}


Figure~\ref{prog:intro-primes} illustrates the use of
a \textit{non-\Iid} loop and conditioning as follows: Repeatedly flip
a coin with bias $p$, counting the number of heads until landing one
tails. Finally, observe that the number of heads counted is a prime
number. What, then, is the posterior over the number of heads $h$? The
true posterior over prime $h$ is given by the probability mass
function (pmf): $\mathsf{Pr}(X = h \: | \: h \text{ is prime})
= \frac{(1 - p)^{h+1}}{\sum_{k \in \mathcal P}{(1 - p)^{k+1}}}$, where
$\mathcal P$ denotes the set of prime
numbers. Table~\ref{table:geometric-primes} shows accuracy and entropy
statistics of the corresponding sampler compiled by \zar.

\begin{table}[t]
  \centering
  \setlength{\tabcolsep}{3pt}
  \begin{tabular}{c c c c c c c c}

    $p$ & $\mu_h$ & $\sigma_h$ & TV & KL & SMAPE & $\mu_{bit}$ &
    $\sigma_{bit}$ \\

    \hline

    $1/2$ & $2.64$ & $1.10$ & $\num{2.33e-3}$ & $\num{6.40e-5}$ &
    $\num{7.63e-2}$ & $9.66$ & $7.21$ \\

    $2/3$ & $3.24$ & $1.93$ & $\num{2.48e-3}$ & $\num{1.10e-4}$ &
    $\num{4.12e-2}$ & $25.31$ & $20.59$ \\

    $1/5$ & $2.19$ & $0.44$ & $\num{7.44e-4}$ & $\num{5.0e-6}$ &
    $\num{5.19e-3}$ & $142.51$ & $132.70$

  \end{tabular}

  \caption{Accuracy and entropy usage for
  Prog.~\ref{prog:intro-primes} with $p = \frac{1}{2}$, $\frac{2}{3}$,
  and $\frac{1}{5}$. $\mu_{bit}$ and $\sigma_{bit}$ are high when $p
  = \frac{1}{5}$ due to low probability of `$h$ is prime',
  illustrating a general weakness (entropy waste) of our rejection
  samplers when conditioning on low-probability events.}

  \label{table:geometric-primes}
\end{table}

\subsection{Uniform Sampling}
\label{subsec:uniform-sampling}

Prog.~\ref{prog:n-sided-die} illustrates a program for rolling an
n-sided die.  Table~\ref{table:n-sided-die} shows accuracy and entropy
usage for $n = 6$, $200$, and
$10000$. Appendix~\ref{app:compare-fldr-optas} contains a comparison
of Ocaml and Python variants of an n-sided die built with Zar with
similar FLDR~\cite{saad2020fldr} and OPTAS~\cite{saad2020sampling}
uniform samplers.

\begin{table}[t]
  \centering
  \setlength{\tabcolsep}{3pt}
  \begin{tabular}{c c c c c c c c}

    $n$ & $\mu_h$ & $\sigma_h$ & TV & KL & SMAPE & $\mu_{bit}$ &
    $\sigma_{bit}$ \\

    \hline

    $6$ & $3.49$ & $1.71$ & $\num{3.86e-3}$ & $\num{5.80e-5}$ &
    $\num{3.87e-3}$ & $3.66$ & $1.33$ \\

    $200$ & $100.42$ & $57.65$ & $\num{1.77e-2}$ & $\num{1.36e-3}$ &
    $\num{1.77e-2}$ & $9.01$ & $2.18$ \\

    $10$k & $5011.87$ & $2892.0$ & $\num{1.24e-1}$ & $\num{7.33e-2}$ &
    $\num{1.28e-1}$ & $15.62$ & $2.74$

  \end{tabular}

  \caption{Accuracy and entropy usage for Prog.~\ref{prog:n-sided-die}
  with $n=6$, $200$, and $10$k (with Shannon entropies $2.59$, $7.64$,
  and $13.29$, respectively). $\mu_{bit}$ and $\sigma_{bit}$ therefore
  show relatively good performance with near entropy-optimality.}

  \label{table:n-sided-die}
\end{table}

\paragraph*{\zar and TensorFlow 2}

We provide
a \href{https://github.com/bagnalla/zar/tree/release-pldi23}{Python 3
package} (built from extracted samplers using
$\mathsf{pythonlib}$~\cite{janestreet2022pythonlib}) exposing a simple
interface for generating samples from verified uniform samplers. To
demonstrate \zar's use as a high-assurance replacement for unverified
samplers, we implement
a \href{https://github.com/bagnalla/zar/tree/release-pldi23/python/tf}{TensorFlow
2~\cite{raschka2019python} project}
(\href{https://github.com/bagnalla/zar/tree/release-pldi23/python/tf}{/python/tf/}
in the source directory) for training an
MNIST~\cite{lecun1998gradient} classifier via stochastic gradient
descent. We observe a negligible effect on training performance and
excellent accuracy on the test set, as expected.

\subsection{Hare and Tortoise}
\label{subsec:hare-and-tortoise}

Our final example shown in Figure~\ref{prog:hare-and-tortoise}
illustrates the use of the discrete Gaussian subroutine (detailed in
the extended version of this
paper~\cite{https://doi.org/10.48550/arxiv.2211.06747}) along with a
non-\iid loop and conditioning to simulate a race between a hare and a
tortoise along a one-dimensional line, and the use of \zar to perform
Bayesian inference~\cite{box2011bayesian}. The tortoise begins with
uniformly-distributed head start $t_0$ and proceeds at a steady pace
of $1$ unit per time step. The hare begins at position $0$ and
occasionally (with probability $\frac{2}{5}$) leaps forward a
Gaussian-distributed distance. The race ends when the hare reaches the
tortoise, and then the terminal state is conditioned on predicate
$P$. For example, by setting $P(time) = time \ge 10$ and querying the
posterior over $t_0$, we ask: ``Given that it took at least 10 time
steps for the hare to reach the tortoise, what are likely values for
the tortoise's head start?'' (see
Figure~\ref{table:hare-and-tortoise}).

\begin{figure}[t]
  \centering
  \begin{subfigure}[t]{.50\textwidth}
    \centering
    \begin{tabular}{c}
    \begin{lstlisting}[language=cpGCL,mathescape=true]
hare_tortoise ($P : \Sigma \rightarrow \prop$) :=
  uniform $10$ ($\lambda n.$ $t_0$ $\leftarrow$ $n$);
  $tortoise$ $\leftarrow$ $t_0$; $hare$ $\leftarrow$ $0$; $time$ $\leftarrow$ $0$;
  while $hare < tortoise$ do
    $time$ $\leftarrow$ $time + 1$;
    $tortoise$ $\rightarrow$ $tortoise + 1$;
    { gaussian $jump$ $4$ $2$;
      $hare$ $\leftarrow$ $hare + jump$ } [$\frac{2}{5}$] { skip }
  end;
  observe $P$
    \end{lstlisting}
    \end{tabular}

    \caption{$\cpGCL$ program simulating a race between
    a \link{hare}{\cwphareztortoise}{hare and tortoise}.}

    \label{prog:hare-and-tortoise}
  \end{subfigure}\hfill%
  \begin{subfigure}[t]{.47\textwidth}
    \setlength{\tabcolsep}{3pt}
    \begin{tabular}{c c c c c}
    $P$ & $\mu_{t_0}$ & $\sigma_{t_0}$ & $\mu_{bit}$ &
    $\sigma_{bit}$ \\
    \hline
    $\true$ & $4.49$ & $2.87$ & $193.88$ & $220.06$ \\
    $time \le 10$ & $3.80$ & $2.79$ & $273.87$ & $378.82$ \\
    $time \ge 10$ & $6.18$ & $2.31$ & $596.68$ & $359.85$ \\
    $time \ge 20$ & $6.40$ & $2.25$ & $1376.74$ & $930.20$ 
    \end{tabular}

    \caption{Accuracy and entropy usage for
    Figure~\ref{prog:hare-and-tortoise}. $\mu_{t_0}$ and $\sigma_{t_0}$
    denote the mean and std deviation of the posterior over the
    tortoise's head start $t_0$, conditioned on $P$. }

  \label{table:hare-and-tortoise}
  \end{subfigure}

  \caption{Hare and tortoise $\cpGCL$ program (left) with accuracy and
  entropy statistics (right).}

  \label{fig:two-thirds}
\end{figure}

\section{Related Work}
\label{sec:related}



\paragraph*{Compilation}
\cite{holtzen2019symbolic, holtzen2020scaling} compile
discrete probabilistic programs with bounded loops and conditioning to
a symbolic representation based on binary decision diagrams
(BDDs)~\cite{darwiche2002knowledge, akers1978binary}, exploiting
independence of variables for efficient exact inference. Our CF trees
are not as highly optimized (and we currently do not support exact
inference), but we remark that BDDs, representing \textit{finite}
Boolean functions, are fundamentally insufficient for programs with
unbounded loops for which no upper bound can be placed on the number
of input bits per sample.

\cite{HuangTM17} compile PPs with continuous
distributions (but not loops) to MCMC samplers for efficient
approximate inference. MCMC algorithms generally provide better
inference performance than \zar (which employs an ``ordinary Monte
Carlo'' (OMC) strategy), but suffer from reliability issues (see
Section~\ref{sec:introduction}). \zar is, to our knowledge, the
only \textit{formally verified} compiler for PPs with loops and
conditioning.

\paragraph*{Verified Probabilistic Systems}

\cite{wang2021sound} implement a type system based on \textit{guide types}
to guarantee compatibility between model and guide functions in a PPL
that compiles to Pyro~\cite{bingham2019pyro}. Pyro is more versatile
than \zar, as it supports continuous distributions and programmable
inference, but provides no formal guarantees on correctness of
compilation or inference. \cite{selsam2018formal} implement Certigrad,
a PP system for stochastic optimization with correctness guarantees in
Lean~\cite{moura2015lean}, but which does not support conditioning or
inference.

\paragraph*{The Conditional Probabilistic Guarded Command Language}
The $\cpGCL$ and its corresponding $\cwp$ semantics were introduced
by~\cite{olmedo2018conditioning} and further developed
by~\cite{kaminski2019advanced} (including discussion of nondeterminism
and expected running time) and~\cite{SzymcakK20} (adding support for
continuous distributions). These works focus on using $\cwp$ as a
program logic for verifying individual programs and metatheoretical
properties of $\cpGCL$, in contrast to \zar which focuses on
verification of compilation to executable samplers.

\paragraph*{Interaction Trees}
Interaction trees have been used to verify compilation of an
imperative programming language~\cite{xia2019interaction}, networked
servers~\cite{koh2019c, letan2020freespec}, an HTTP key-value
server~\cite{ZhangHK0LXBMPZ21}, and transactional
objects~\cite{LesaniXKBCPZ22}. The \zar system presents a novel
application of interaction trees to verified executable semantics of
probabilistic programs, and employs a novel domain-theoretic framework
for reasoning about them (Section ~\ref{subsec:generating-itrees}).

\paragraph*{Evaluation of PPLs}
\cite{dutta2018testing} implement a testing framework for
PPLs called ProbFuzz that generates random test programs for various
PPLs and compares their inference results to detect irregularities. We
expect that \zar can be incorporated into ProbFuzz as a reference
against which other discrete PPLs should be evaluated.

\section{Conclusion}
\label{sec:conclusion}

This paper presents the first formalization of $\cpGCL$ and its $\cwp$
semantics in a proof assistant, and implements \zar, the first
formally verified compiler from a discrete PPL to proved-correct
executable samplers. \zar uses a novel intermediate representation, CF
trees, to optimize and debias probabilistic choices. CF trees are
compiled to executable interaction trees encoding the sampling
semantics of source programs in the random bit model. The full
compilation pipeline is formally proved to satisfy an equidistribution
theorem showing that the empirical distribution of generated samples
converges to the true posterior distribution of the source $\cpGCL$
program. \zar's backend supports extraction to OCaml and has been used
to generate samplers for a collection of probabilistic programs
including the discrete Gaussian distribution.

\smallskip
\noindent\textbf{Acknowledgments.}
We thank the anonymous reviewers and the paper's shepherd, Jan Hoffmann, for their comments. Banerjee's research was based on work supported by the National Science Foundation (NSF), while working at the Foundation. Any opinions, findings, and conclusions or recommendations expressed in this article are those of the authors and do not necessarily reflect the views of the NSF.

\smallskip
\noindent\textbf{Data Availability Statement.} Sources for Zar
and all examples in this paper are available on Zenodo with the
identifier \href{https://doi.org/10.5281/zenodo.7809333}{10.5281/zenodo.7809333}~\cite{bagnall_alexander_2023_7809333}.

\appendix 

\section{Debiasing CF trees}
\label{app:debias}

CF trees generated by the \zar compiler may have arbitrary $p \in
[0,1]$ bias values at choice nodes. To obtain sampling procedures in
the random bit model, a bias-elimination transformation is applied to
CF trees, replacing all probabilistic choices by fair coin-flipping
schemes that implement the same behavior. The CF trees resulting from
this transformation have $p=\frac{1}{2}$ at all choice
nodes. Figure~\ref{fig:debias2} shows an example of the debiasing
transformation applied to a $\cons{Choice}{}$ node with $p
= \frac{2}{3}$.

\begin{figure}[H]
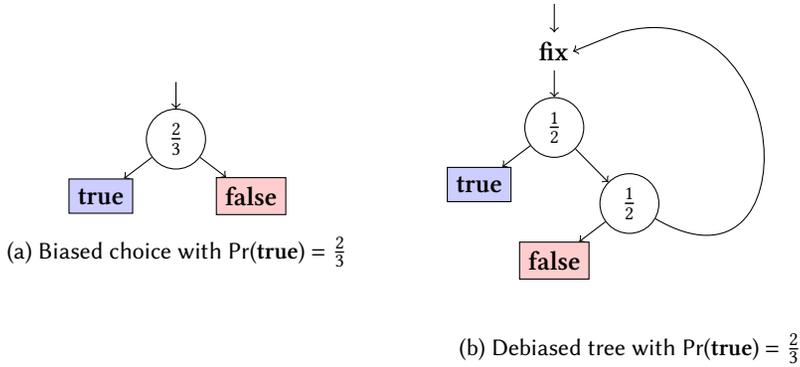

  \centering
  \begin{subfigure}{.5\textwidth}
    \centering
    \ctikzfig{biased2}
    \caption{Biased choice with Pr($\true$) $= \frac{2}{3}$}
    \label{fig:two-thirds-choice2}
  \end{subfigure}%
  \begin{subfigure}{.5\textwidth}
    \centering
    \ctikzfig{debiased2}
    \caption{Debiased tree with Pr($\true$) $= \frac{2}{3}$}
    \label{fig:two-thirds-debiased2}
  \end{subfigure}

  \caption{$\cons{Choice}{}$ CF tree with $Pr(\true) = \frac{2}{3}$
  (left) and corresponding debiased CF tree (right).}

  \label{fig:debias2}
\end{figure}

The algorithm for translating a `$\cons{choice}{} \: p \: k$' node
with rational bias $p = \dfrac{n}{d}$ and subtrees $t_1 = k \: \true$
and $t_2 = k \: \false$ goes as follows:
\begin{enumerate}
  \item Recursively translate $t_1$ and $t_2$, yielding $t_1'$ and
    $t_2'$ respectively,
  \item choose $m : \nat$ such that $2^{m-1} < d \leq 2^m$,
  \item generate a perfect CF tree of depth $m$ with all terminal
  nodes marked by a special $\cons{loopback}{}$ value,
  \item replace the first $n$ terminals with copies of subtree $t_1'$,
    and the next $d$ terminals with copies of subtree $t_2'$, leaving
    $s^m - n - d$ $\cons{loopback}{}$ nodes remaining,
  \item coalesce duplicate leaf nodes to eliminate redundancy,
  \item wrap the tree in a $\cons{fix}{}$ constructor with guard
  condition that evaluates to true on the $\cons{loopback}{}$ value,
  and
  \item replace the original $\cons{choice}{}$ node with the newly
  generated tree.
\end{enumerate}

In essence, the biased choice is replaced by a rejection sampler that
simulates a biased coin by repeated flips of a fair one. An
implementation of the choice translation algorithm is in the file
`uniform.v' under the name `$\mathsf{bernoulli\_tree}$'. The overall
debiasing transformation is then a straightforward recursive traversal
of the input CF tree, using $\mathsf{bernoulli\_tree}$ to replace
biased $\cons{Choice}{}$ nodes with equivalent subtrees containing
only unbiased choices.

\begin{definition}[\link{debias}{\cwpdebias}{$\mathsf{debias}$}]
  \label{def:debias}
  
  For $t : \cftree$ a CF tree, define $\mathsf{debias} \: t : \cftree$ inductively on $t$:
  \begin{center}
    \setlength{\tabcolsep}{3pt}
    \begin{tabular}{l l l}
      \rowcolor{lightgray}
      \multicolumn{3}{l}{$\link{debias}{\cwpdebias}{\mathsf{debias}}
      : \cftree \rightarrow \cftree$} \\
      \hline
      $\cons{Leaf}{} \: \sigma$ & $\triangleq$ &
      $\cons{Leaf}{} \: \sigma$ \\
      $\cons{Fail}{}$ & $\triangleq$ & $\cons{Fail}{}$ \\
      $\cons{Choice}{} \: p \: k$ & $\triangleq$ &
      $\mathsf{bernoulli\_tree \: p \bind{} \lambda b. \: \text{if }
      b \text{ then } \mathsf{debias} \: (k \: \true) \text{ else
      } \mathsf{debias} \: (k \: \false)}$ \\
      $\cons{Fix}{} \: \sigma \: e \: g \: k$ & $\triangleq$ &
      $\cons{Fix}{} \: \sigma \: e \: (\mathsf{debias} \circ g) \:
      (\mathsf{debias} \circ k)$
    \end{tabular}
  \end{center}
\end{definition}

The essential results for the debiasing transformation now follow:
\textsf{debias} preserves tcwp semantics and produces trees in which
all choices are unbiased.

\begin{theorem}[\link{debias}{\cwptcwpzdebias}{$\mathsf{debias}$ preserves $\tcwp$ semantics}]
  \label{theorem:debias-preserves-tcwp}

  Let $t : \cftree$ be a CF tree and $f : \expectation$ an
  expectation. Then, $\tcwp \: (\mathsf{debias} \: t) \: f = \tcwp \:
  t \: f. \qed$
\end{theorem}

\begin{theorem}[\link{debias}{\cwptreezunbiasedzdebias}{$\mathsf{debias}$ produces unbiased trees}]
  \label{theorem:debias-unbiased}

  Let $t : \cftree$ be a CF tree. Then, $p = \frac{1}{2}$ for every
  $\: \cons{Choice}{} \: p \: k$ in $t$. $\qed$
\end{theorem}

\section{Comparison with FLDR and OPTAS}
\label{app:compare-fldr-optas}

The Fast Loaded Dice Roller (FLDR)~\cite{saad2020fldr} is a time- and
space-efficient algorithm for rolling an $n$-sided die, with
implementations available in Python and C. Related to FLDR is
OPTAS~\cite{saad2020sampling}, a system for optimal approximate
sampling from discrete distributions wrt.~a user-specified number of
random bits, also with implementations in Python and
C. Table~\ref{table:comparison-samplers} shows a comparison of a
$200$-sided die in FLDR and OPTAS (with $32$-bit precision and the
``hellinger'' kernel) with OCaml and Python implementations of the
$200$-sided die based on a variant of
$\link{uniform_Z}{\cwpuniformztree}{\mathsf{uniform\_tree}}$ from
Section~\ref{subsec:compiling-to-cf-trees} that uses binary-encoded
integers rather than unary natural numbers. Initialization time is
negligible for both \zar and FLDR.


\begin{table}[H]
  \centering
  \setlength{\tabcolsep}{2pt}
  \begin{tabular}{l c c c c c c c c c}

    & $\mu_x$ & $\sigma_x$ & TV & KL & SMAPE & $\mu_{bit}$ &
    $\sigma_{bit}$ & $T_{init}$ & $T_s$ \\

    \hline

    \zar \small{(OCaml)} & $99.43$ & $57.73$ & $\num{1.91e-2}$ &
    $\num{1.75e-3}$ & $\num{2.41e-2}$ & $9.0$ & $2.16$ & $<\!\!1$ms &
    $105$ms \\

    \zar \small{(Py)} & $99.87$ & $57.63$ & $\num{1.95e-2}$ &
    $\num{1.03e-3}$ & $\num{2.20e-2}$ & $9.01$ & $2.19$ & $<\!\!1$ms &
    $292$ms \\

    FLDR \small{(C)} & $99.39$ & $57.79$ & $\num{1.96e-2}$ & $\num{1.18e-3}$ &
    $\num{2.21e-2}$ & $9.01$ & $2.18$ & $<\!\!1$ms & $6$ms \\

    FLDR \small{(Py)} & $99.32$ & $57.70$ & $\num{2.08e-2}$ &
    $\num{1.36e-3}$ & $\num{2.33e-2}$ & $9.0$ & $2.16$ & $<\!\!1$ms &
    $290$ms \\

    OPTAS \small{(C)} & $99.50$ & $57.74$ & $\num{1.85e-2}$ & $\num{1.20e-3}$
    & $\num{2.10e-2}$ & $8.55$ & $1.27$ & $3$ms & $5$ms \\

    OPTAS \small{(Py)} & $99.58$ & $57.69$ & $\num{2.12e-2}$ &
    $\num{1.37e-3}$ & $\num{2.37e-2}$ & $8.55$ & $1.27$ & $15$ms &
    $330$ms \\

  \end{tabular}

  \caption{Comparison of $200$-sided die samplers with output
  $x$. $T_{init}$ denotes time elapsed over construction and
  initializion of the sampler, and $T_s$ the total time to generate
  $100$k samples.}

  \label{table:comparison-samplers}
\end{table}

\section{Discrete Gaussian}
\label{app:discrete-gaussian}

We define discrete variants of Laplace and Gaussian distributions
(based on~\cite{canonne2020discrete}) as reusable subroutines for
larger $\cpGCL$ programs (e.g., the hare and tortoise program in
Section~\ref{subsec:hare-and-tortoise}). These subroutines differ from
$\mathsf{flip}$ in Def.~\ref{def:flip} by making use of local
variables. Although $\cpGCL$ lacks built-in support for procedure
calls (which can be done in principle, as
in~\cite{olmedo2016reasoning}), they can be shallowly embedded if we
take careful account of variables modified (i.e, ``clobbered'') within
subroutines.

\subsection{Inverse Exponential Bernoulli}

To sample from a discrete Laplace, we first require a subroutine for
sampling from a Bernoulli distribution with inverse exponential
bias. We begin with a preliminary routine
($\mathsf{bernoulli\_exponential\_0\_1}$) for the special case of
$\mathbf{0} \le \gamma \le \mathbf{1}$, which modifies variables $k$
and $a$ (used as a counter and loop flag, respectively), followed by
its generalization ($\mathsf{bernoulli\_exponential}$) to
$\mathbf{0} \le \gamma$, additionally modifying variables $i$ and $b$
(also a counter and loop flag).

\begin{table}[t]
  \centering
  \begin{tabular}{c}
    \begin{lstlisting}[language=cpGCL,mathescape=true]
bernoulli_exponential_0_1 ($out$ : ident) ($\gamma : \Sigma \rightarrow \mathbb{Q}$) :=
  $k$ $\leftarrow$ $0$; $a$ $\leftarrow$ $\true$;
  while $a$ do { $k$ $\leftarrow$ $k + 1$ } [$\frac{\gamma}{k + 1}$] { $a$ $\leftarrow$ $\false$} end;
  if even $k$ then $out$ $\leftarrow$ $\true$ else $out$ $\leftarrow$ $\false$ end
    \end{lstlisting}
  \end{tabular}

  \caption{\link{gaussian}{\cwpbernoullizexponentialzqzw}{Sampling
  from Bernoulli(exp($-\gamma$))}, where
  $\mathbf{0} \le \gamma \le \mathbf{1}$}

  \label{prog:bernoulli-exponential-0-1}
\end{table}

\begin{figure}[t]
  \centering
  \begin{tabular}{c}
    \begin{lstlisting}[language=cpGCL,mathescape=true]
bernoulli_exponential ($out$ : ident) ($\gamma : \Sigma \rightarrow \mathbb{Q}$) :=
  if $\gamma \le 1$
  then bernoulli_exponential_0_1 $out$ $\gamma$
  else $i$ $\leftarrow$ $1$; $b$ $\leftarrow$ $\true$;
      while $b \land i \le \gamma$ do bernoulli_exponential_0_1 $b$ $\mathbf{1}$; $i$ $\leftarrow$ $i + 1$ end;
      if $b$ then bernoulli_exponential_0_1 $out$ ($\gamma - \lfloor \gamma \rfloor$) else $out$ $\leftarrow$ $0$ end
    \end{lstlisting}
  \end{tabular}

  \caption{\link{gaussian}{\cwpbernoullizexponential}{Sampling from
  Bernoulli(exp($-\gamma$))}, where $\mathbf{0} \le \gamma$}

  \label{prog:bernoulli-exponential}
\end{figure}

\begin{table}[t]
  \centering
  \setlength{\tabcolsep}{3pt}
  \begin{tabular}{c c c c c c c c}

    $\gamma$ & $\mu_{out}$ & $\sigma_{out}$ & TV & KL & SMAPE &
    $\mu_{bit}$ & $\sigma_{bit}$ \\

    \hline

    $1/2$ & $0.61$ & $0.49$ & $\num{1.86e-3}$ & $\num{1.0e-5}$ &
    $\num{1.95e-3}$ & $2.54$ & $2.16$ \\

    $3/2$ & $0.23$ & $0.42$ & $\num{1.36e-3}$ & $\num{8.0e-6}$ &
    $\num{1.96e-3}$ & $3.84$ & $3.59$ \\

    $10$ & $\num{9.0e-5}$ & $\num{9.49e-3}$ & $\num{4.50e-5}$ &
    $\num{2.50e-5}$ & $\num{1.65e-1}$ & $4.56$ & $5.11$

  \end{tabular}

  \caption{Accuracy and entropy usage for
  Figure~\ref{prog:bernoulli-exponential}.}

  \label{table:bernoulli-exponential}
\end{table}

\subsection{Discrete Laplace}
A discrete analogue
$\mathsf{Lap}_{\mathbb{Z}}(b)$~\cite{canonne2020discrete} of the
Laplace distribution (useful for, e.g., differential
privacy~\cite{ghosh2009universally}, and as a subroutine for the
discrete Gaussian in the following section) with scale parameter $b$
is defined by the probability mass function
$\mathsf{Pr}_{\mathsf{Lap}_{\mathbb{Z}}(b)}(X = x) = \frac{e^{1/b} -
1}{e^{1/b} + 1} \cdot e^{-|x|/b}$. Figure~\ref{prog:discrete-laplace}
shows a $\cpGCL$ program for sampling from
$\mathsf{Lap}_\mathbb{Z}(\frac{t}{s})$ for positive integers $s$ and
$t$.


\begin{figure}[t]
  \centering
  \begin{tabular}{c}
    \begin{lstlisting}[language=cpGCL,mathescape=true]
laplace ($out$ : ident) ($s \: t : \nat$) :=
  $lp$ $\leftarrow$ $\true$;
  while $lp$ do
    uniform $t$ ($\lambda u.$
       bernoulli_exponential $d$ ($\lambda s.$ $\frac{u}{t}$);
       if $d$ then
         $v$ $\leftarrow$ $0$; bernoulli_exponential $il$ $1$;
         while $il$ do $v$ $\leftarrow$ $v + 1$; bernoulli_exponential $il$ $1$ end;
           $x$ $\leftarrow$ $u + t \cdot v$; $y$ $\leftarrow$ $\frac{x}{s}$; flip $c$ $\frac{1}{2}$;
           if $c \land y = 0$ then skip
           else $lp$ $\leftarrow$ $\false$; $out$ $\leftarrow$ $(1 - 2 \cdot [c]) \cdot y$
       else skip)
  end
    \end{lstlisting}
  \end{tabular}

  \caption{\link{gaussian}{\cwplaplace}{Sampling from
  $\mathsf{Lap}_{\mathbb{Z}}$}. Modified variables: $k$, $a$, $i$,
  $b$, $lp$, $d$, $v$, $il$, $x$, $y$, and $c$. Variables $lp$ and
  $il$ (``loop'' and ``inner loop'') are used for control
  flow. See~\cite{canonne2020discrete} for explanation and
  proof-of-correctness of the sampling algorithm.}

  \label{prog:discrete-laplace}
\end{figure}

\begin{table}[t]
  \centering
  \setlength{\tabcolsep}{3pt}
  \begin{tabular}{c c c c c c c c}

    $s, t$ & $\mu_{out}$ & $\sigma_{out}$ & TV & KL & SMAPE &
    $\mu_{bit}$ & $\sigma_{bit}$ \\

    \hline

    $1, 2$ & $\num{1.79e-2}$ & $2.81$ & $\num{3.51e-3}$ &
    $\num{4.20e-4}$ & $\num{1.64e-1}$ & $10.47$ & $7.04$ \\

    $2, 1$ & $\num{1.79e-3}$ & $0.60$ & $\num{1.47e-3}$ &
    $\num{7.10e-5}$ & $\num{5.30e-2}$ & $9.77$ & $8.17$ \\

    $5, 2$ & $\num{-8.50e-4}$ & $0.44$ & $\num{1.24e-3}$ &
    $\num{1.09e-4}$ & $\num{1.37e-1}$ & $15.53$ & $12.38$

  \end{tabular}

  \caption{Accuracy and entropy usage for
  Figure~\ref{prog:discrete-laplace} with scale parameter
  $\frac{t}{s}$.}

  \label{table:discrete-laplace}
\end{table}

\subsection{Discrete Gaussian}

A discrete analogue $\mathcal
N_{\mathbb{Z}}(\mu, \sigma^2)$~\cite{canonne2020discrete} of the
Gaussian (``normal'') distribution (useful for, e.g., lattice-based
cryptography~\cite{zhao2020cosac}, and as a subroutine for the
hare-and-tortoise program in Section~\ref{subsec:hare-and-tortoise})
with parameters $\mu$ and $\sigma$ is defined by the probability mass
function $\mathsf{Pr}_{\mathcal N_{\mathbb{Z}}(\mu, \sigma^2)}(X = x)
= \frac{e^{-(x - \mu)^2 / 2 \sigma^2}}{\sum_{y : \mathbb{Z}}{e^{-(y
- \mu)^2 / 2 \sigma^2}}}$. Figure~\ref{prog:discrete-gaussian} shows a
$\cpGCL$ program for sampling from $\mathcal
N_{\mathbb{Z}}(\mu, \sigma^2)$.


\begin{figure}[t]
  \centering
  \begin{tabular}{c}
    \begin{lstlisting}[language=cpGCL,mathescape=true]
gaussian_0 ($z$ : ident) ($\sigma : \mathbb{Q}$) :=
  $ol$ $\leftarrow$ $\false$;
  while $\lnot ol$ do laplace $z$ $1$ $\lfloor \sigma \rfloor + 1$; bernoulli_exponential $ol$ ($\lambda s.$ $\frac{(|z| - \frac{\sigma^2}{t})^2}{2 \sigma^2}$) end

gaussian ($out$ : ident) ($\mu : \Sigma \rightarrow \mathbb{Z}$) ($\sigma : \mathbb{Q}$) :=
  gaussian $out$ $\sigma$; $out$ $\leftarrow$ $out + \mu$
    \end{lstlisting}
  \end{tabular}

  \caption{\link{gaussian}{\cwpgaussian}{Sampling from $\mathcal
  N_{\mathbb{Z}}(\mu, \sigma^2)$}. Note that the entropy usage depends
  only on $\sigma$ and not $\mu$. Modified variables: $k$, $a$, $i$,
  $b$, $lp$, $d$, $v$, $il$, $x$, $y$, $c$, $ol$, $z$. Variable $ol$
  (``outer loop'') is used for control
  flow. See~\cite{canonne2020discrete} for explanation and
  proof-of-correctness of the sampling algorithm.}

  \label{prog:discrete-gaussian}
\end{figure}

\begin{table}[t]
  \centering
  \setlength{\tabcolsep}{3pt}
  \begin{tabular}{c c c c c c c c}

    $\mu, \sigma$ & $\mu_z$ & $\sigma_z$ & TV & KL & SMAPE &
    $\mu_{bit}$ & $\sigma_{bit}$ \\

    \hline

    $0, 1$ & $\num{-3.03e-3}$ & $1.0$ & $\num{2.71e-3}$ &
    $\num{1.03e-4}$ & $\num{4.49e-2}$ & $26.68$ & $24.43$ \\

    $10, 2$ & $10.0$ & $2.0$ & $\num{3.69e-3}$ & $\num{1.16e-4}$ &
    $\num{7.22e-2}$ & $37.61$ & $29.10$ \\

    $-50, 5$ & $-50.01$ & $5.01$ & $\num{6.11e-3}$ & $\num{4.46e-4}$ &
    $\num{5.70e-2}$ & $43.66$ & $31.20$

  \end{tabular}

  \caption{Accuracy and entropy usage for
  Figure~\ref{prog:discrete-gaussian} with mean $\mu$ and variance
  $\sigma^2$.}

  \label{table:discrete-gaussian}
\end{table}



\bibliography{main}

\end{document}

%% file: listingsconfig.tex
\usepackage{listings}
\lstdefinelanguage{myC}{
  language=C,
  basicstyle=\linespread{1.2}\ttfamily,
  showspaces=false,              
  showstringspaces=false,        
  showtabs=false,    
  tabsize=2,                      
  captionpos=b,                   
  breaklines=true,                
  breakatwhitespace=false,
  escapeinside={\%*}{*)},        
  keywordstyle=\bfseries\color{black},    
  numberstyle=\tiny\color{gray},
}
\lstdefinelanguage{myinlineC}{
  language=myC,
  basicstyle=\ttfamily
}
\lstdefinelanguage[x86gasm]{Assembler}[x86masm]{Assembler}{%
,basicstyle=\ttfamily\singlespacing
,morekeywords={rax,rbx,rcx,rdx,rip,rdi,rsi,rsp,subq,decl,movq
              ,movl,xorl,imull,popq,popl,pushl}%
,morekeywords=[2]{.file,.section,.string,.text,.globl,.cfi_startproc
                 ,.cfi_def_cfa_offset,.cfi_endproc,.size,.ident}%
}
\definecolor{cmmtcolor}{named}{OliveGreen}
\lstdefinelanguage{Coq}{
,morekeywords={match,end,Definition,Inductive,Lemma,Theorem,Record,
               Hypothesis,Variable,Section,End,case,of,if,then,else,
               is,let,in,do,return,with,Extract,Constant,Inlined,Inline,
               Extraction,Fixpoint,Program,Function,Class,CoInductive,
               CoFixpoint,Variant,Instance,Context}
,morecomment=[s]{(*}{*)}
,keywordstyle=\bfseries\color{MidnightBlue}
,commentstyle={\color{cmmtcolor}}
,basicstyle=\linespread{1.1}\sffamily
,columns=fullflexible
,numberstyle=\tiny\color{gray}
,escapeinside={@}{@}
,literate=
    {:=}{{$\triangleq\;$}}1
    {<-}{{$\leftarrow\;$}}1
    {=>}{{$\Rightarrow\;$}}1
    {->}{{$\rightarrow\;$}}1
    {<->}{{$\leftrightarrow\;$}}1
    {<==}{{$\leq\;$}}1
    {\\/}{{$\vee\;$}}1
    {/\\}{{$\land\;$}}1
    {ffun}{{$\mathsf{ffun}$}}1    
    {fun}{{$\lambda$}}1
    {forall}{{$\forall$}}1
    {exists}{{$\exists$}}1
    {Z}{{$\mathbb{Z}$}}1
    {Z0}{{$\mathbb{Z}_0$}}1
    {<=}{{$\leq\;$}}1
    {>=}{{$\geq\;$}}1
    {<>}{{$\neq\;$}}1                
}

\lstdefinelanguage{PPL}{
,morekeywords={if,then,else,return,observe,while}
,keywordstyle=\bfseries\color{MidnightBlue}
,commentstyle={\color{cmmtcolor}}
,basicstyle=\linespread{1.2}\sffamily
,columns=fullflexible
,numberstyle=\tiny\color{gray}
,escapeinside={@}{@}
,literate=
    {<-}{{$\leftarrow\;$}}1
}

\lstdefinelanguage{cpGCL}{
,morekeywords={if,then,else,observe,while,do,end,skip,uniform}
,keywordstyle=\bfseries\color{MidnightBlue}
,morecomment=[s]{(*}{*)}
,commentstyle={\color{cmmtcolor}}
,basicstyle=\linespread{1.0}\sffamily
,columns=fullflexible
,numberstyle=\tiny\color{gray}
,escapeinside={@}{@}
,literate=
    {<-}{{$\leftarrow\;$}}1
}

\lstdefinelanguage{grammar}{
,morekeywords={skip,abort,observe,ite,while}
,keywordstyle=\bfseries\color{MidnightBlue}
,commentstyle={\color{cmmtcolor}}
,basicstyle=\linespread{1.0}\ttfamily
,numberstyle=\tiny\color{gray}
,escapeinside={@}{@}
}

%% file: db.tex
\newcommand{\cwpcpGCL}{82}
\newcommand{\cwpCSkip}{83}

\newcommand{\cwpCAssign}{85}
\newcommand{\cwpCSeq}{86}
\newcommand{\cwpCIte}{87}
\newcommand{\cwpCChoice}{88}
\newcommand{\cwpCUniform}{89}
\newcommand{\cwpCWhile}{90}
\newcommand{\cwpCObserve}{91}

\newcommand{\cwpcwp}{168}
\newcommand{\cwpwpz}{130}
\newcommand{\cwpwlpz}{149}
\newcommand{\cwptree}{23}
\newcommand{\cwptreezbind}{32}
\newcommand{\cwpLeaf}{24}
\newcommand{\cwpFail}{25}
\newcommand{\cwpChoice}{26}
\newcommand{\cwpFix}{27}

\newcommand{\cwptcwp}{53}
\newcommand{\cwptwpz}{21}
\newcommand{\cwptwlpz}{37}

\newcommand{\cwpcompile}{16}
\newcommand{\cwpcwpztcwp}{117}
\newcommand{\cwpco}{99}
\newcommand{\cwpaCPO}{49}
\newcommand{\cwpcotreezext}{278}
\newcommand{\cwpitwp}{100}
\newcommand{\cwpitreezsampler}{43}

\newcommand{\cwptiezitree}{206}

\newcommand{\cwpcpGCLztozitree}{402}
\newcommand{\cwpitreezpreimage}{103}
\newcommand{\cwpcwpzitwp}{465}

\newcommand{\cwptozitreezopen}{169}

\newcommand{\cwpGetBool}{41}
\newcommand{\cwptwpzuniformztreezsum}{1146}
\newcommand{\cwpuniformztree}{403}
\newcommand{\cwptcwpzdebias}{125}
\newcommand{\cwptreezunbiasedzdebias}{134}
\newcommand{\cwpdebias}{25}

\newcommand{\cwpuniform}{71}
\newcommand{\cwpcpGCLzsampleszequidistributed}{519}
\newcommand{\cwpcotreezpreimage}{80}

\newcommand{\cwpflip}{31}
\newcommand{\cwpduelingzcoins}{21}

\newcommand{\cwpdie}{21}

\newcommand{\cwpbernoullizexponentialzqzw}{34}
\newcommand{\cwpbernoullizexponential}{62}
\newcommand{\cwplaplace}{127}
\newcommand{\cwpgaussian}{214}
\newcommand{\cwphareztortoise}{82}

%% file: tree.tikzstyles
\tikzstyle{square}=[fill={rgb,255: red,205; green,205; blue,255}, draw=black, shape=rectangle]
\tikzstyle{circle}=[fill=white, draw=black, shape=circle]
\tikzstyle{grey_square}=[fill={rgb,255: red,229; green,229; blue,229}, draw=black, shape=rectangle]
\tikzstyle{red_square}=[fill={rgb,255: red,255; green,205; blue,205}, draw=black, shape=rectangle]
\tikzstyle{big box}=[fill=white, draw=black, shape=rectangle, minimum width=8.5cm, minimum height=4cm]
\tikzstyle{light grey square}=[fill={rgb,255: red,245; green,245; blue,245}, draw=black, shape=rectangle]

\tikzstyle{arrow}=[->, fill=none]
\tikzstyle{reverse arrow}=[<-]

%% file: biased2.tikz
\begin{tikzpicture}
	\begin{pgfonlayer}{nodelayer}
		\node [style=circle] (0) at (-2.25, 1.75) {$\frac{2}{3}$};
		\node [style=square] (1) at (-4.25, 0.25) {$\true$};
		\node [style={red_square}] (3) at (-0.25, 0.25) {$\false$};
		\node [style=none] (9) at (-2.25, 3.25) {};
	\end{pgfonlayer}
	\begin{pgfonlayer}{edgelayer}
		\draw [style=arrow] (0) to (1);
		\draw [style=arrow] (9.center) to (0);
		\draw [style=arrow] (0) to (3);
	\end{pgfonlayer}
\end{tikzpicture}

%% file: debiased2.tikz
\begin{tikzpicture}
	\begin{pgfonlayer}{nodelayer}
		\node [style=circle] (0) at (-2.25, 3.75) {$\frac{1}{2}$};
		\node [style=square] (1) at (-4.25, 2.25) {$\true$};
		\node [style=circle] (2) at (-0.25, 1.75) {$\frac{1}{2}$};
		\node [style={red_square}] (3) at (-2.25, 0.25) {$\false$};
		\node [style=none] (9) at (-2.25, 5.25) {};
		\node [style=none] (10) at (-0.5, 6.25) {};
		\node [style=none] (11) at (-2.25, 5.75) {$\cons{fix}{}$};
		\node [style=none] (12) at (-2.25, 7) {};
		\node [style=none] (13) at (-2.25, 6.25) {};
		\node [style=none] (14) at (-1.75, 5.75) {};
	\end{pgfonlayer}
	\begin{pgfonlayer}{edgelayer}
		\draw [style=arrow] (0) to (1);
		\draw [style=arrow] (9.center) to (0);
		\draw [style=arrow] (2) to (3);
		\draw [style=arrow] (0) to (2);
		\draw [in=15, out=-30, looseness=2.50] (2) to (10.center);
		\draw [style=arrow] (12.center) to (13.center);
		\draw [style=arrow] (10.center) to (14.center);
	\end{pgfonlayer}
\end{tikzpicture}

%% file: two_thirds_bias_intervals_small.pdf_tex
\begingroup%
  \makeatletter%
  \providecommand\color[2][]{%
    \errmessage{(Inkscape) Color is used for the text in Inkscape, but the package 'color.sty' is not loaded}%
    \renewcommand\color[2][]{}%
  }%
  \providecommand\transparent[1]{%
    \errmessage{(Inkscape) Transparency is used (non-zero) for the text in Inkscape, but the package 'transparent.sty' is not loaded}%
    \renewcommand\transparent[1]{}%
  }%
  \providecommand\rotatebox[2]{#2}%
  \newcommand*\fsize{\dimexpr\f@size pt\relax}%
  \newcommand*\lineheight[1]{\fontsize{\fsize}{#1\fsize}\selectfont}%
  \ifx\svgwidth\undefined%
    \setlength{\unitlength}{559.37594677bp}%
    \ifx\svgscale\undefined%
      \relax%
    \else%
      \setlength{\unitlength}{\unitlength * \real{\svgscale}}%
    \fi%
  \else%
    \setlength{\unitlength}{\svgwidth}%
  \fi%
  \global\let\svgwidth\undefined%
  \global\let\svgscale\undefined%
  \makeatother%
  \begin{picture}(1,0.2514095)%
    \lineheight{1}%
    \setlength\tabcolsep{0pt}%
    \put(0,0){\includegraphics[width=\unitlength,page=1]{two_thirds_bias_intervals_small.pdf}}%
    \put(0.24116778,0.10697147){\color[rgb]{0,0.78431373,0}\makebox(0,0)[lt]{\lineheight{1.25}\smash{\begin{tabular}[t]{l}{\color{black} $\true$}\end{tabular}}}}%
    \put(0,0){\includegraphics[width=\unitlength,page=2]{two_thirds_bias_intervals_small.pdf}}%
    \put(0.24876153,0.05883299){\color[rgb]{0,0,0}\makebox(0,0)[lt]{\begin{minipage}{0.11234393\unitlength}\raggedright \end{minipage}}}%
  \end{picture}%
\endgroup%